\documentclass[pra, aps, onecolumn, groupedaddress, showpacs, superscriptaddress]{revtex4}
\usepackage{amsmath}
\usepackage{amsfonts}
\usepackage{amssymb}
\usepackage{graphicx}
\usepackage{hyperref}
\usepackage{bm}
\usepackage{color}
\hypersetup{backref,
colorlinks=true,
linkcolor=cyan,
linktoc=page,
citecolor=cyan,
urlcolor=cyan}
\usepackage{subfigure}

\newcommand{\pr}[1]{\ensuremath{\left[#1\right]}} 
\newcommand{\pc}[1]{\ensuremath{\left(#1\right)}}

\newcommand{\ket}[1]{\ensuremath{\left\vert#1\right\rangle}} 
\newcommand{\md}[1]{\ensuremath{\left\vert#1\right\vert}} 
\newcommand{\av}[1]{\ensuremath{\left\langle#1\right\rangle}}

\begin{document}
\title{Cooling and entanglement of multimode graphene resonators via vacuum fluctuations}
\author{Sofia Ribeiro}
\affiliation{Instituto de Telecomunica\c{c}\~oes, Lisbon, Portugal}
\email{sofia.ribeiro@lx.it.pt}
\author{Hugo Ter\c{c}as}
\affiliation{Instituto de Plasmas e Fus\~ao Nuclear, Lisbon, Portugal}


\begin{abstract}
Sympathetic laser cooling of a single mode graphene membrane coupled to an atomic cloud interacting via Casimir-Polder forces has been recently proposed. Here, we extend this study to the effect of secondary graphene membrane whose frequency may be far or close to resonance. We show that if the two mechanical modes are close together, it is possible to simultaneously cool both modes. Conversely, if the two frequencies are set far apart, the secondary mode does not affect the cooling of the first one. We also study the entanglement properties of the steady-state using the logarithmic negativity. We show how stationary mechanical entanglement between two graphene sheets can be generated by means of vacuum fluctuations. Moreover, we find that, within feasible experimental parameters, large steady-state acoustomechanical entanglement, i.e. entanglement between the phononic and mechanical mode, $E_N \approx 5$, can be generated.
\end{abstract}
\maketitle

\section{Introduction}

In the technological push towards miniaturization, one of the ultimate goals is to build nanomechanical resonators that are only one atom thick. Two-dimensional (2D) nanoresonators offer a unique platform for quantum technologies thanks to their low mass, low stress, and high quality factors. Graphene's extraordinary electronic and optical properties hold great promise for applications in photonics, electronics and optomechanical systems \cite{RMP81_109_2009}. Lately, a great effort has been put in harnessing the mechanical properties of graphene for mass sensing \cite{NatNano4_861_2009}, studying nonlinear mechanics \cite{NatNano6_339_2011,NanoLett12_198_2012}, and voltage tunable oscillators \cite{NanoLett12_4681_2012,NatNanotech8_923_2013}. Building these miniaturized mechanical systems evolved to the idea of devising small structures based on graphene, where neutral atoms and graphene are held in close proximity. Such a hybrid system would consist of atoms that are manipulated by laser light, and graphene sheets that could, for instance, be controlled by electrical currents or piezoelectrics \cite{wang2007,liu2016}. 

Atom-graphene coupling can be achieved via vacuum forces \cite{acta2008}. Casimir-Polder interaction is a promising tool to manipulate and cool quantum states of mechanical oscillators. Proposals range from shielding vacuum fluctuations with graphene sheets \cite{ShieldingPaper}, quantum sensing of graphene motion \cite{PRL112_223601_2014}, manipulation of the atomic states to create ripples on demand \cite{RipplesPaper}, and passive sympathetic cooling carbon nanotubes by immersing them in cold atom clouds \cite{PRA88_043623_2013}. In a recent study, the authors proposed a method to actively sympathetic cool graphene membranes by laser-cooling an atomic cloud placed at a few $\mu$m distance \cite{CoolingPaper}. This overcomes the important limitation of radiation-pressure cooling of graphene, a process that is known to be hindered by its broad absorption spectrum \cite{stauber2008}. The influence of vacuum forces in optomechanical systems has been the focus of many recent studies. In reference \cite{PRA88_063849_2013}, W.~Nie and co-authors studied the effect of the Casimir force between a dielectric nanosphere coupled to a movable mirror of a Fabry-P\'erot cavity. They have shown that the ground state cooling of the nanosphere is achieved for certain sphere-mirror distances and that it can be optimized by tuning the mirror oscillation frequency. In reference~\cite{SciChinPhysMA57_2276_2014}, the study is extended to consider two nanospheres trapped near the cavity mirrors by an external driving laser. By tuning the external control parameters and the cavity-sphere distance, they found to be possible to achieve large steady-state optomechanical entanglement. In reference~\cite{OptExpress23_30970_2015}, the authors focused on the Casimir-Polder interaction of an ensemble of quantum emitters coupled to a movable mirror inside a cavity. It is shown that vacuum forces not only greatly enhance the effective damping rate but also lead, in the bad cavity limit, to the ground-state cooling of the mechanical motion. 

\begin{figure}[t!]
\centering
\includegraphics[width=12cm]{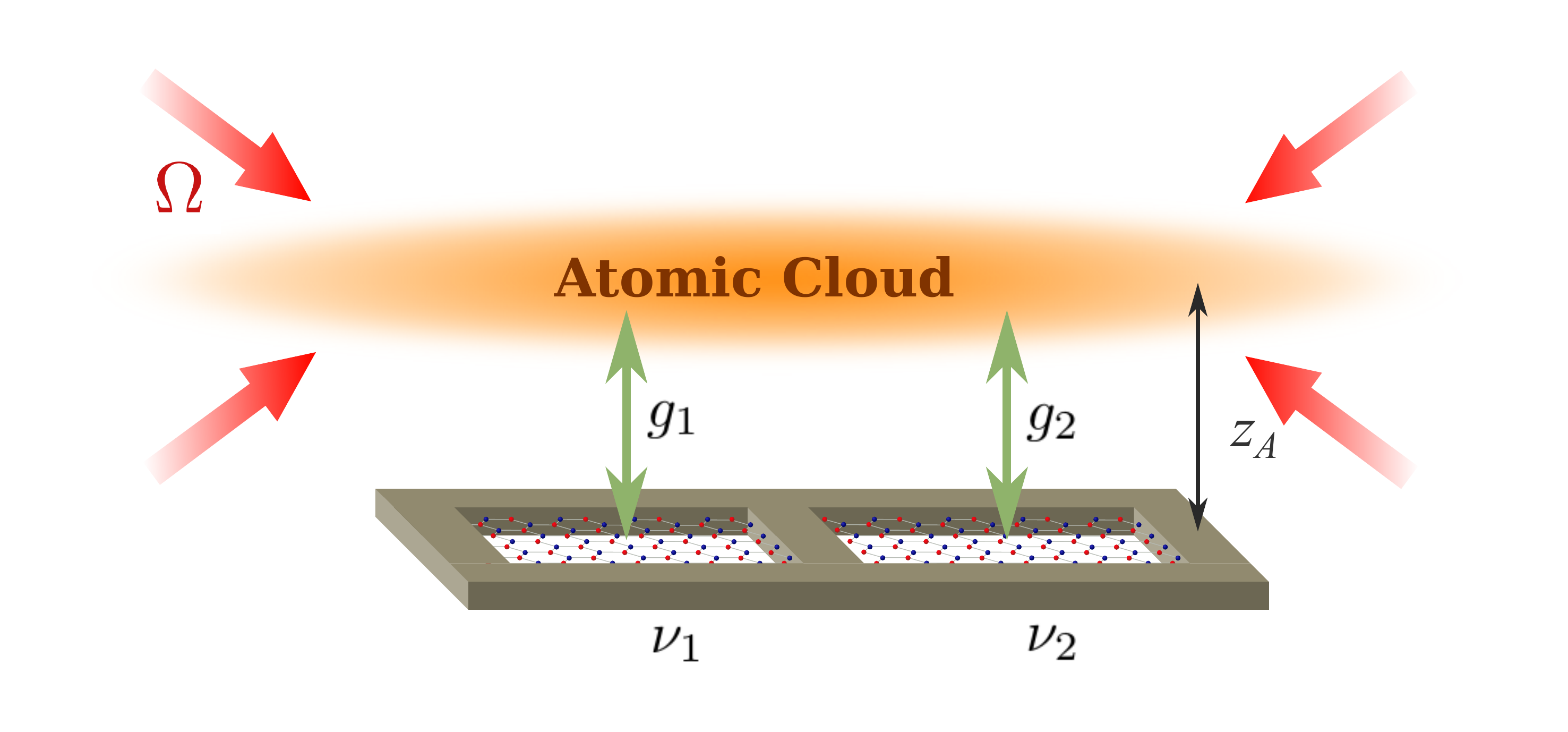}
\caption{(Colour online) Schematic representation of the experimental setup. A cold atomic gas is placed at a distance $z_A$ from a system with two graphene membranes suspended on a substrate. Quantum excitations in the atomic cloud (phonons) are coupled to the flexural (out-of-plane) modes of graphene via vacuum forces. The cooling and entanglement of the phonons of the gas can be done with the help of the cooling laser with Rabi frequency $\Omega$. For our calculations, we have chosen an atomic cloud of $^{87}$Rb.\label{Fig:setup}} 
\end{figure}

Accessing and controlling the quantum ground-state is a milestone in all optomechanical system, as it allows us to harness the quantum behaviour of a macroscopic object and to explore the quantum-classical boundary. However, up to now, most theoretical treatments of cooling have focused on a \textit{single phononic mode -- single mechanical mode} interaction. Here, we shall extend the treatment of cooling introduced in reference~\cite{CoolingPaper} by including the effect of secondary modes whose resonance frequency is not far from that of the mechanical mode of interest. This secondary mode can be considered on a single membrane device or in a multi-mode membrane device, as illustrated in figure~\ref{Fig:setup}. In addition, we will also study the steady-state entanglement between the different mechanical modes. Entanglement is a typical property of the quantum world, non existent in the classical realm. However, there is nothing in the quantum mechanical principles that prevents macroscopic systems to be entangled. In fact, entanglement has been experimentally achieved in microscopic quantum systems such as photons \cite{PRL75_4337_1995,Nat430_54_2004,NatPhys3_91_2007,NatPhot6_225_2012}, ions \cite{Nat459_683_2009}, electrons \cite{PRL117_050402_2016}, buckyballs \cite{Nat401_680_1999,AmJPhys71_4_2003}, and in macroscopic systems such as diamonds \cite{Science334_1253_2011}. Beside the inherent fundamental interest, the ability to create entangled states as also useful applications, as in high precision and metrology applications where entangled states represent a very sensitive probe \cite{PRD19_1669_1979} and can have profound implications to optical information science and quantum computing \cite{PRA86_042306_2012,PRL110_253601_2013}.

In this work, we investigate the macroscopic mechanical entanglement generation in two or more coupled graphene nanoresonators. As we are about to show, entanglement is achieved via vacuum forces that couple the different membranes through the elementary excitations (phonons) of an cold atomic cloud. We propose experimentally feasible schemes to create and probe acoustomechanical entanglement,
i.e. entanglement between the phonon mode (here playing the role of a cavity photon) and the graphene flexural mode.

This paper is organized as follows. In Sec.~\ref{sec:theory}, the theoretical model of the optomechanical system with multi-modes flexurons coupled via electromagnetic vacuum fluctuations to a single atomic cloud is introduced.  In Sec.~\ref{sec:simultaneouscooling}, we study the simultaneous cooling of flexuron modes in near by membranes. We extend our study also to the acoustomechanical and mechanical entanglement in different configurations. Finally, in Sec.~\ref{sec:conclusions}, we provide some concluding remarks. 

\section{Theoretical model for the experimental setup \label{sec:theory}}

\subsection{Description of vibrating graphene\label{sec:vibrations}}

Although classic elastic theory focus on very large systems, it has been shown that the elastic continuum theory is still valid for a small graphene flake \cite{PRB64_235406_2001}. Due to thermal fluctuations, the graphene sheet undergoes mechanical out-of-plane vibrations (flexural phonons), which can be well described within the Kirchhoff-plate theory of elasticity \cite{PRB88_115418_2013}. In what follows, we will consider a squared graphene flake with both edges clamped to a substrate subject to a restoring force. 
Having determined the eigenmodes of the graphene membrane, the Hamiltonian for the flexural modes easily follows from the canonical quantization of the dynamics of the out-of-plane vibrations. Thus, we write the Hamiltonian as
\begin{eqnarray}
\hat{H}_\text{F} &= \frac{1}{2} \int d^2 r \; \left[  D \nabla^4 \mathbf{u}  \pc{\mathbf{r},t} + h \ddot{ \mathbf{u} } \pc{\mathbf{r},t} 
+ 2 t_{\rm cl} \nabla^2 \mathbf{u} \pc{\mathbf{r},t}  \right].
\end{eqnarray}
We then express the out-of-plane displacement in the form
\begin{eqnarray}
\hat{\mathbf{u}} \pc{\mathbf{r}} = \frac{1}{\sqrt{2}} \sum_{\mathbf{k},\sigma} \phi_{\mathbf{k},\sigma} \pc{\mathbf{r}} \mathbf{e}_\sigma \pc{\hat{f}_{\mathbf{k}, \sigma} + \hat{f}^\dagger_{\mathbf{k},\sigma}}
\end{eqnarray}
with two polarizations $\sigma=(x,y)$ and satisfying the normalization condition $\av{\phi_\mathbf{k}, \phi_{\mathbf{k}'}} = \hbar / \pc{M \nu_k} \delta_{\mathbf{kk}'}$, where $M$ is the membrane mass and $\nu_k$ the vibration frequency of the flexural mode $\nu = \sqrt{\frac{D}{\rho} k^4 + \frac{2 t_{\rm cl}}{\rho}k^2 }$, with $\rho$ areal mass density, $D = \frac{1}{12} Y h^3 / \pc{1-\upsilon^2}$ the bending modulus, $Y\sim 1$~TPa the Young modulus, $\upsilon = 0.17$ the Poisson ratio, $h = 3.35$~\AA~the thickness of the plate and $t_{\rm cl}$ the clamping tension (for simplicity, we consider it to be equal along both $x$ and $y$ directions). 
The flexural operators obey the usual bosonic commutation relation $\pr{\hat{f}_{\mathbf{k}, \sigma}, \hat{f}^\dagger_{\mathbf{k}',\sigma'}} = \delta_{\mathbf{kk}'} \delta_{\sigma,\sigma'}$. Thus, the Hamiltonian of the vibrating graphene is simply given as follows 
\begin{eqnarray}
\hat{H}_\text{flex} = \sum_{\mathbf{k}, \sigma} \hbar \nu_{\mathbf{k}, \sigma}  \hat{f}^\dagger_{\mathbf{k}, \sigma} \hat{f}_{\mathbf{k},\sigma}.
\end{eqnarray}

\subsection{Description of the total Hamiltonian}

As described above, our system of interest consists of a laser-cooled two-dimensional cloud of atoms that is placed a few micrometres from one or more graphene membranes (see figure~\ref{Fig:setup}). Due to the tight confinement in the perpendicular direction, the (transverse) phonon modes of the atomic cloud are quantized. The initial Hamiltonian contains five terms: i) the energy of the electronic states of the atoms; ii) the energy associated to the quantized atomic motion; iii) the quantized modes of the membranes; iv) the coupling between the laser and the atomic motion and v) Casimir-Polder (CP) interaction between the atoms and the membrane (for a detailed derivation, please refer to reference\cite{CoolingPaper}). We first proceed to the adiabatic elimination of the excited electronic states and assume that the atoms are cooled enough to be in the Lamb-Dicke regime. This means that we are in a situation where the difference between the atomic phonon modes is much larger than the difference of the flexural modes in graphene, such that we can safely retain the lowest phononic mode only. Finally, we obtain the effective Hamiltonian, which is nonlinear in the phonon operator $\hat{a}$, and the effective Lindblad operator as \cite{CoolingPaper}
\begin{eqnarray}
&\hat{H}_\text{eff} =
\hbar \omega \hat{a}^\dagger \hat{a}
 + \hbar \sum_j \nu_j \hat{f}^\dagger_j \hat{f}_j + i \hbar \sum_j g_j \hat{a}^\dagger \hat{a} \pc{\hat{f}_j +\hat{f}^\dagger_j}+ i \hbar \xi \pc{\hat{a}^\dagger + \hat{a} }  \\
& \mathcal{L}_\text{eff} \pc{\hat{O}} = \gamma \pc{2 \hat{a}^\dagger \hat{O} \hat{a} - \hat{O}\hat{a}^\dagger \hat{a} - \hat{a}^\dagger \hat{a} \hat{O}}.
\end{eqnarray}
Here, we have defined the reduced quantities 
\begin{eqnarray*}
\omega = \omega_\text{ph} - \frac{\eta^2 \hbar \Omega^2 \Delta}{4 \Delta^2 + \Gamma^2} + \sum_j \omega^{\text{CP}}_j, \quad \quad
 \xi =\frac{\eta \Omega^2 \Delta }{4 \Delta^2 + \Gamma^2}, \quad \quad 
 \gamma =\frac{\Gamma  \eta^2 \Omega^2 }{2 \pc{\Gamma +4 \Delta ^2}},
\end{eqnarray*}
where $\omega_\text{ph}$ is the energy of the phonon excitation, $\eta$ the Lamb-Dicke parameter, $\Delta$ the detuning between the laser and the electronic transition, $\Gamma$ the atomic spontaneous emission rate and $\Omega$ the Rabi frequency. The coupling strength between the graphene sheet and the atomic cloud is given by \cite{HugoSofia2015}
\begin{eqnarray}
g_j= 2 q_j\sqrt{ \frac{\hbar}{2 m \nu_j}} n_0 \, \omega^{\text{CP}}_j
\end{eqnarray}
with $n_0$ being the atomic density. In the non-retarded limit, that is, when the atom-surface distance $z_A$ is small when compared to the effective atomic transition wavelength $z_A \ll c / \omega_\text{eg} $, the Casimir-Polder potential becomes $U_{\text{CP}} = C_3 / z_A^3$. For Rubidium atom near a graphene sheet, one finds $C_3 = −215.65$~Hz$\mu$m$^3$ \cite{RubidiumPaper}. After performing a Fourier transformation, the fundamental Casimir-Polder frequency reads \cite{CoolingPaper}
\begin{eqnarray}
 \omega^{\text{CP}}_1 = 2 \pi C_3 \frac{e^{-q_1 z_A}}{z_A}.
\end{eqnarray}

\subsection{Heisenberg-Langevin equations of motion}

An appropriate treatment of the problem requires including other different effects, the main one being the losses in the flexural modes which are quantified by the energy dissipation rate $\kappa_j = \nu_j / Q_j$, where $Q_j$ is the mechanical quality factor. In reference~\cite{NatNano9_820_2014}, the authors demonstrated coupling between a multilayer graphene resonator with quality factors up to 2.2$\times 10^5$, which results in dissipative rates of the orders of a few tens of Hz or lower. Therefore, by defining the dimensionless \textit{position} and \textit{momentum operator} operators 
\begin{eqnarray}
\hat{q}_j = \frac{i \pc{\hat{f}^\dagger_j + \hat{f}_j}}{\sqrt{2}}, \quad
\hat{p}_j = \frac{\hat{f}^\dagger_j - \hat{f}_j}{\sqrt{2}},
\end{eqnarray}
with $\pr{\delta \hat{q}_k, \delta \hat{p}_j}=i \delta_{kj}$, and adopting the formalism of quantum Langevin equations, we find
\begin{align}
\dot{\hat{a}} &= -i \omega \hat{a} - i  \sum_j g_j \sqrt{2} \hat{a} \hat{q}_j + \xi - \gamma \hat{a}, \label{eq:dota}\\
\dot{\hat{q}}_j &= - \nu_j \hat{p}_j, \label{eq:dotq} \\
\dot{\hat{p}}_j &= \nu_j  \hat{q}_j - \kappa_j  \hat{p}_j - \sqrt{2} g_j \hat{a}^\dagger \hat{a} + \zeta, \label{eq:dotp}
\end{align}
where the mechanical modes of graphene are affected by a viscous force with damping rate $\kappa$ and by a Brownian stochastic force with zero mean value $\zeta$, with correlation function \cite{PRA63_023812_2001}
\begin{eqnarray}
\av{\zeta(t) \zeta(t')} = \frac{\kappa_j}{\nu_j} \int \frac{d \omega}{2 \pi} e^{-i \omega (t-t')} \omega
\pr{\coth \pc{\frac{\hbar \omega}{2 k_B T}}+1},
\end{eqnarray}
where $k_B$ is the Boltzmann constant and $T$ is the graphene temperature. $\zeta (t)$ is a Gaussian quantum stochastic process and non-Markovian, i.e., neither its correlation function or its commutator are proportional to a Dirac delta. However, we can simplify the thermal noise contribution. $k_B T / \hbar \sim 10^{11} s^{-1}$ even at cryogenic temperatures \cite{PRL113_027404_2014,NJP10_095009_2008}, as so, it is always much larger than all the other parameters. 
At these higher values of frequency the position spectrum is negligible and therefore one can safely approximate the integral \cite{PRA77_033804_2008}
\begin{eqnarray}
\frac{\kappa_j \omega}{\nu_j} \coth \pc{\frac{\hbar \omega}{2 k_B T}} \simeq \kappa_j \frac{2 k_B T}{\hbar \nu_j} \simeq \kappa_j \pc{2 m_j +1},
\end{eqnarray} 
where $m_j = \pc{\exp \pc{\hbar \nu_j / k_B T} -1}^{-1}$ is the mean thermal number of the mode $j$.

Following ref.~\cite{NJP10_095009_2008,PRA77_033804_2008,PRL98_030405_2007}, we will arrive at a system of linearised quantum Langevin equations (see more details in \ref{App:Linearisation})
\begin{align}
\delta \dot{\hat{q}}_j &= - \nu_j \delta \hat{p}_j, \\
\delta \dot{\hat{p}}_j &=  \nu_j  \delta \hat{q}_j - \kappa_j  \delta \hat{p}_j - 2 g_j \md{\alpha} \delta \hat{X} + \zeta, \\
\delta \dot{\hat{X}} &=   \vartheta_{(N)}\delta \hat{Y} -  \gamma \delta \hat{X},\\
\delta \dot{\hat{Y}} &=  -  \vartheta_{(N)}\delta \hat{X} -  \gamma \delta \hat{Y} + \sum_j 2 \md{\alpha} g_j \delta \hat{q}_j,
\end{align}
where we have defined $\vartheta_{(N)} = \omega + \sum_j 2 g_j^2 \md{\alpha}^2 / \nu_j$, with $N$ being the total number of modes considered.
These can be written in a compact form as
\begin{eqnarray}
\dot{u} (t) = A u(t) + n(t),
\end{eqnarray}
where we defined the fluctuation vector $u(t)$ and the noise vector $n(t)$ and $A$ is the drift matrix that governs the dynamics of the expectation values. Since the dynamics is linearised, the quantum steady-state of fluctuations is a zero-mean multipartite Gaussian state, fully characterized by its correlation matrix $\mathcal{V}$ that can be find by solving
\begin{eqnarray}
A \mathcal{V}+ \mathcal{V} A^T = -D. \label{eq:Lyapunov}
\end{eqnarray}
where $D= \text{Diag}\pr{0,\kappa_j \pc{2 m_j +1 },0,0 }$ is the diagonal matrix determined by the noie correlation function (for the complete derivation please see \ref{App:Linearisation})
This equation is linear for $\mathcal{V}$ and can be straight-forwardly solved. The stationary variances of the mechanical modes are given by the diagonal terms of $\mathcal{V}$, $\av{\delta \hat{q}_1^2} = \mathcal{V}_{11}$, $\av{\delta \hat{p}_1^2} = \mathcal{V}_{22}$, $\av{\delta \hat{q}_2^2} = \mathcal{V}_{33}$, $\av{\delta \hat{p}_2^2} = \mathcal{V}_{44}$,$\dots$, $\av{\delta \hat{X}^2} = \mathcal{V}_{N-1,N-1}$, $\av{\delta \hat{Y}^2} = \mathcal{V}_{NN}$.
\begin{figure}
\centering
\begin{minipage}{.5\textwidth}
  \centering
  \includegraphics[width=7.5cm]{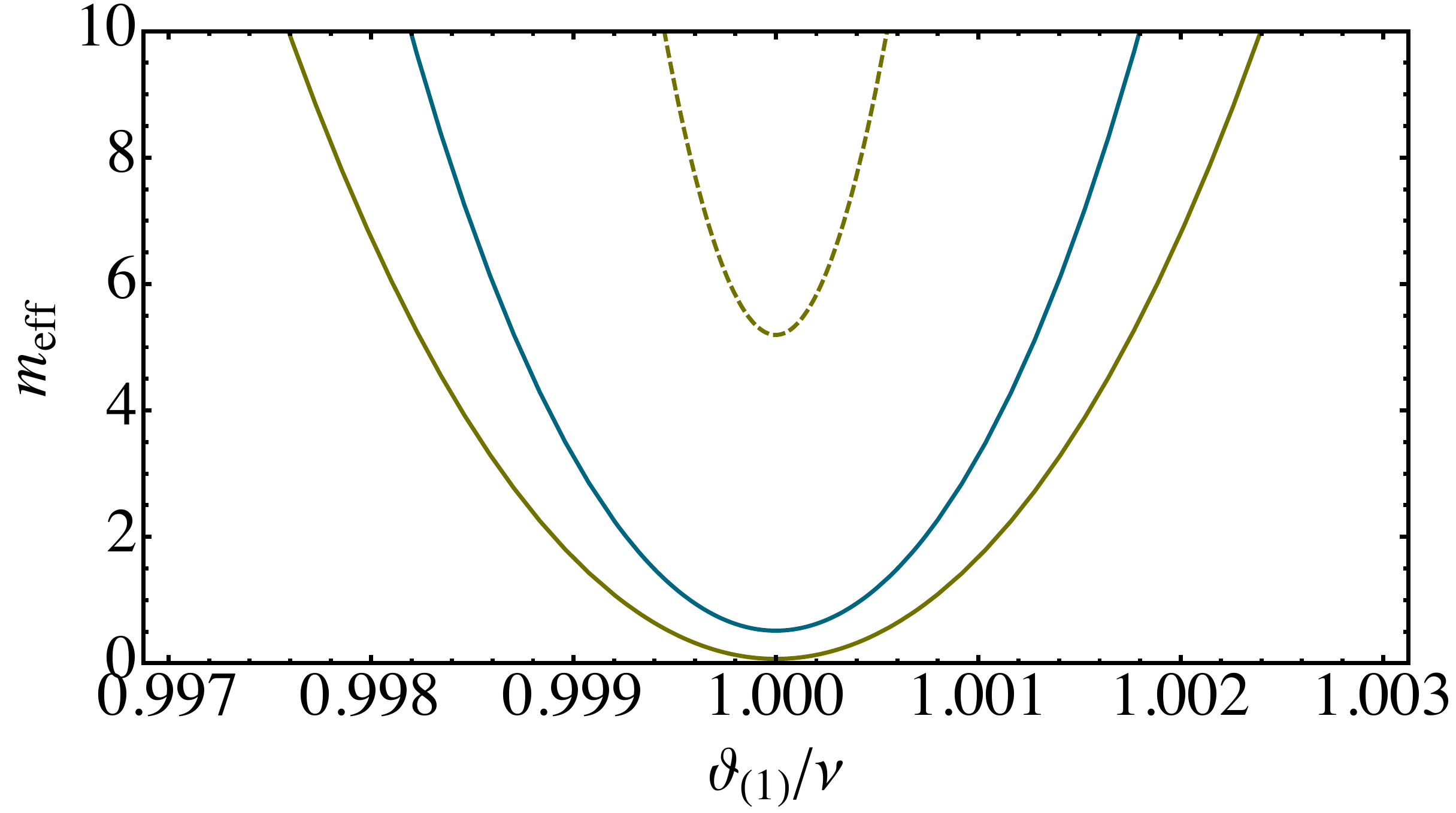}
\end{minipage}%
\begin{minipage}{.5\textwidth}
  \centering
  \includegraphics[width=7.5cm]{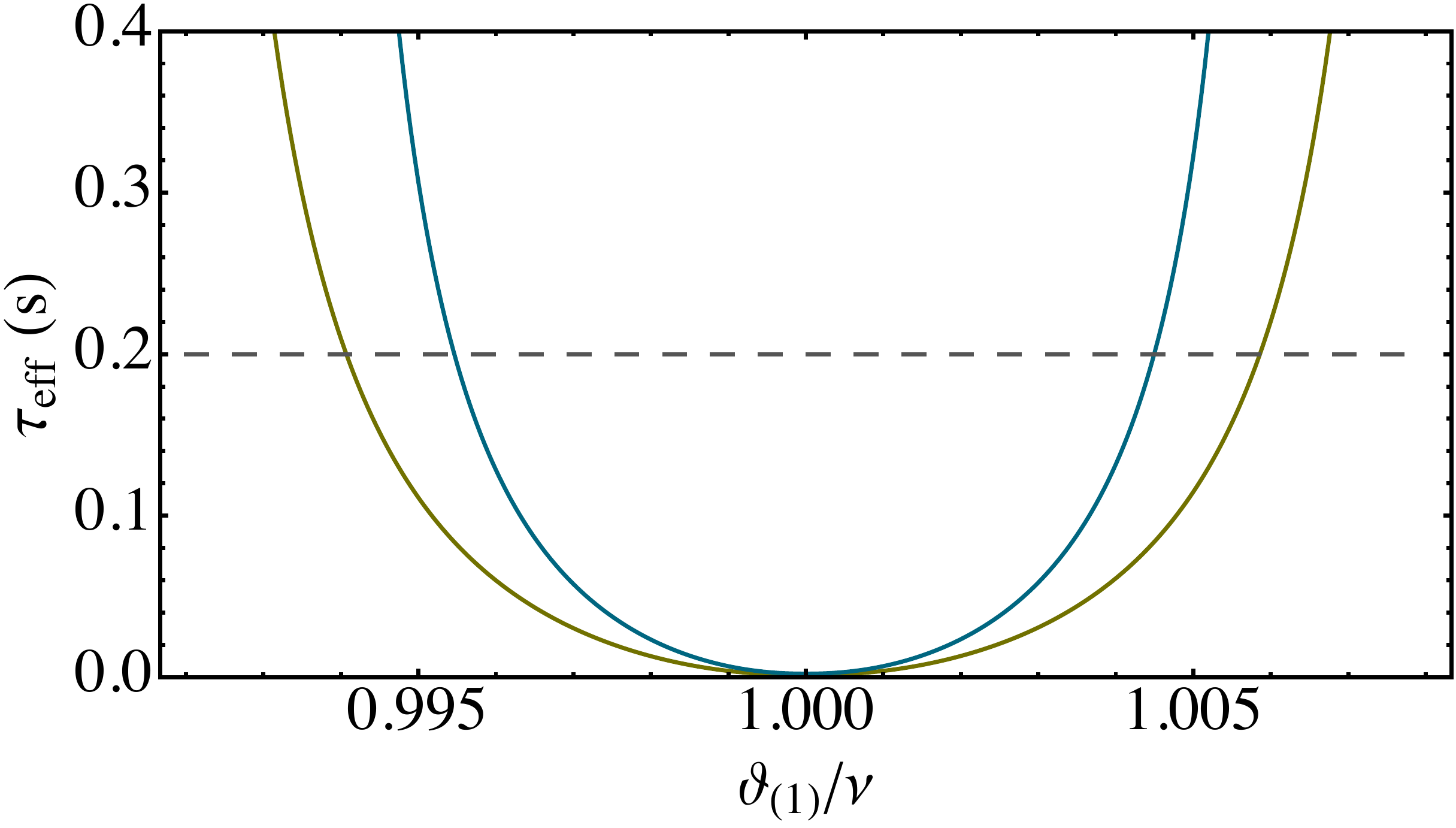}
\end{minipage}
\caption{(Colour online) Steady-state flexural number $m_\text{eff}$ (left) and decay rate of one flexural mode of a single graphene sheet versus normalized detuning $\vartheta_{(1)}/\nu$. The atomic and mechanical parameters are $\Gamma = 6.1$~MHz, $\omega_\text{ph}=477$~Hz, $\Omega = 12$~MHz, $\Delta = 45$~MHz, $\eta =0.15$, $\kappa =2$~Hz and $\nu = 2$~MHz and $T=0.01$~K, that corresponds to an initial occupancy $m=10^2$ . The yellow solid line corresponds to a coupling strength $g=-6.5$~kHz (dashed yellow line corresponds to $T=0.1$~K with initial occupancy $m=10^3$), the blue solid line to $g=-5$~kHz and the dashed gray line to $g=0$. \label{Fig:1mode}}
\end{figure}

\begin{figure}
\centering
\begin{minipage}{.5\textwidth}
  \centering
  \includegraphics[width=7.5cm]{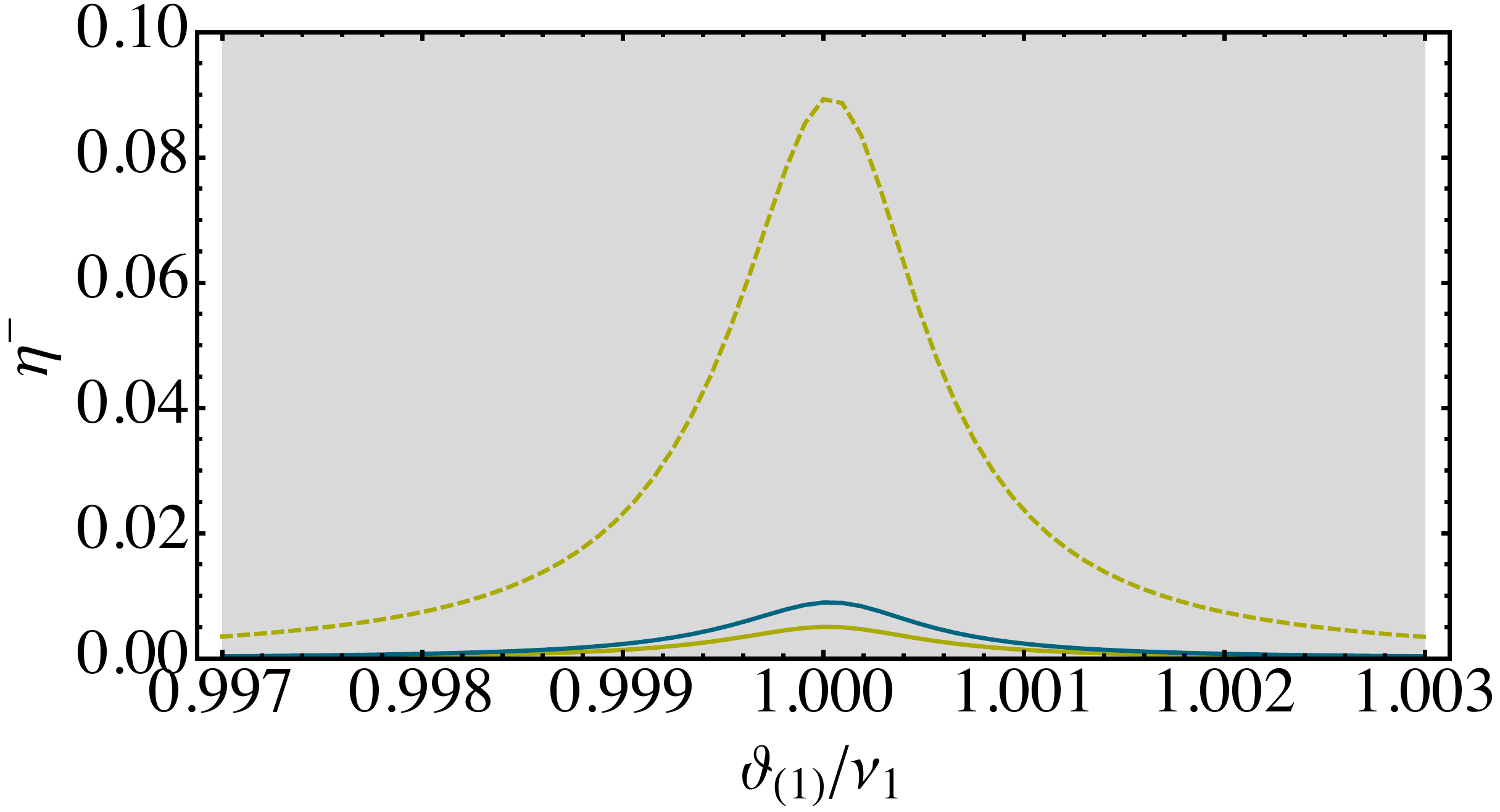}
\end{minipage}%
\begin{minipage}{.5\textwidth}
  \centering
  \includegraphics[width=7.5cm]{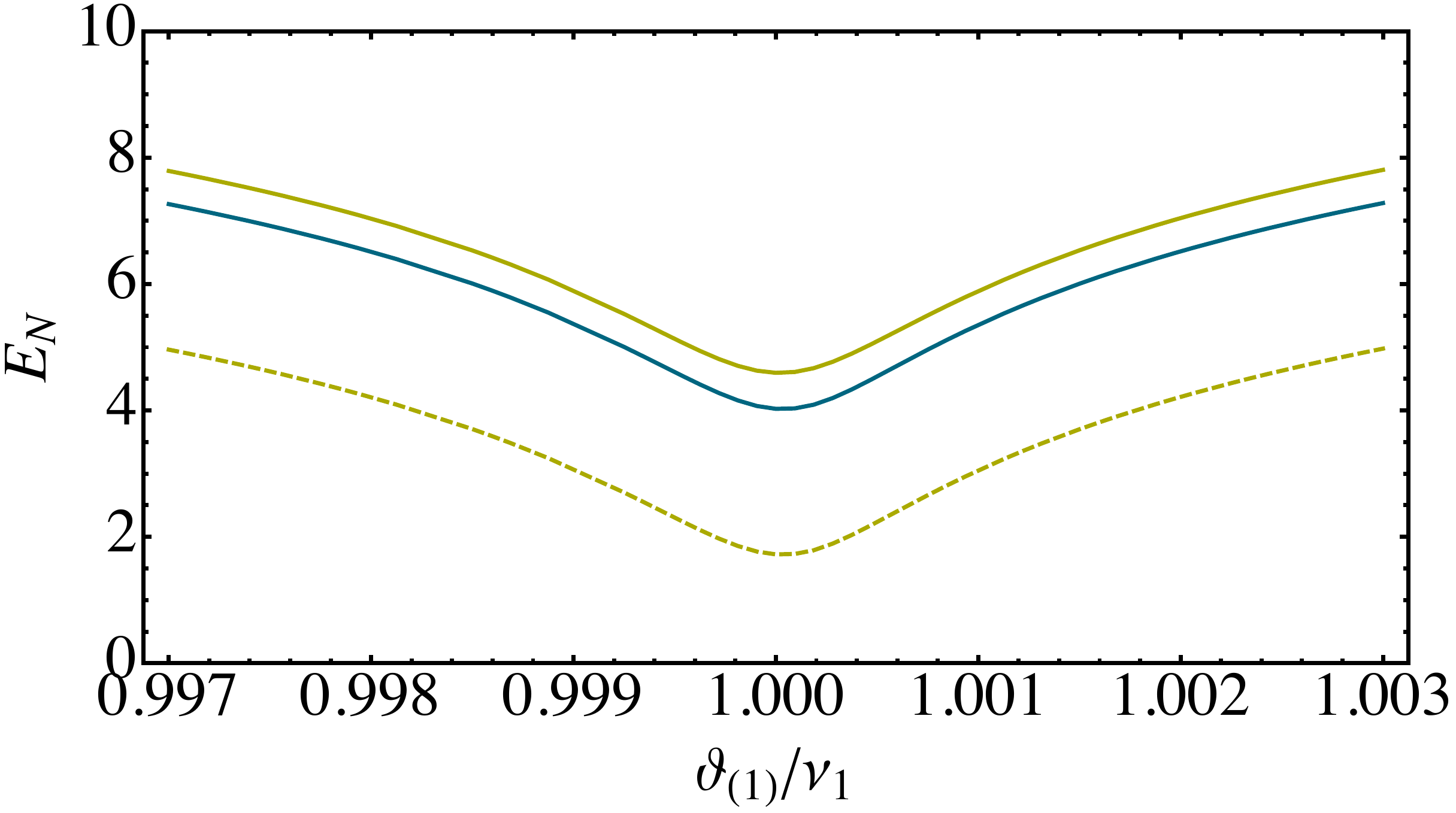}
\end{minipage}
 \caption{(Colour online) Acoustomechanical entanglement $\eta^-$ (left) and logaritmic negativity $E_N$ (right) for a single graphene sheet versus normalized detuning $\vartheta_{(1)}/\nu$. The atomic and mechanical parameters are $\Gamma = 6.1$~MHz, $\omega_\text{ph}=477$~Hz, $\Omega = 12$~MHz, $\Delta = 45$~MHz, $\eta =0.15$, $\kappa =2$~Hz and $\nu = 2$~MHz. The solid lines are for $T=0.01$~K and the dashed line to $T=0.1$~K. The yellow lines correspond to a coupling strength $g=-6.5$~kHz and the blue line to $g=-5$~kHz. \label{Fig:1modeEntanglement}}
\end{figure}
 
%%%%%%%%%%%%%%%%%%%%%%%%%%%%%%%%%%%%%%%%%%%%%%%%%%%%%%%%%%%%%%%%%%
\section{Steady-state and entanglement of multimode graphene \label{sec:simultaneouscooling}}

At the steady-state, the energy of each mechanical mode can be written in the terms of the variances of the corresponding position and momentum operators
\begin{eqnarray}
U_j = \hbar \nu_j m_\text{eff}^j = - \frac{\hbar \nu_j}{2} \pr{ \av{\delta q_j^2} + \av{\delta p_j^2} +1} ,
\end{eqnarray}
where $m_\text{eff}^j = - \pc{ \av{\delta q_j^2} + \av{\delta p_j^2} + 1} / 2$ is the effective occupation number of the $j$th mode.

\subsection{Single mechanical mode}

If only one mechanical mode is considered, the drift matrix assumes the following explicit form
\begin{eqnarray}
A_{(1)} = \begin{pmatrix}
0 & - \nu & 0 & 0 \\
\nu & -\kappa & - 2 \md{\alpha} g & 0 \\
0 & 0 & -\gamma & \vartheta_{(1)} \\
- 2 \md{\alpha} g & 0 & -\vartheta_{(1)} & - \gamma
\end{pmatrix}.
\end{eqnarray}
The stability conditions can be determined by applying the Routh-Hurwitz criterion \cite{PRA35_5288_1987}
\begin{eqnarray}
\nu^2 \pc{\vartheta_{(1)}^2 + \gamma^2} - 4 \nu \md{\alpha}^2 g^2 \vartheta_{(1)} &> 0, \\
2 \gamma \kappa \left[\vartheta_{(1)}^4 + \vartheta_{(1)}^2 \pc{\kappa^2 + 2 \kappa\gamma + 2 \gamma^2 - 2 \nu^2} \right.
\nonumber \\ 
\left.
+ \pc{\kappa \gamma + \gamma^2 + \nu^2}^2 \right]
+4 \nu \md{\alpha}^2 g^2 \vartheta_{(1)} \pc{\kappa + 2 \gamma }^2 &>0 . 
\end{eqnarray}

In agreement with the results of ref.~\cite{CoolingPaper}, with an appropriate choice of $g$ and $\vartheta_{(1)}$, it is possible to have effective steady-states with a low number of flexural modes, (typically $m_\text{eff} < 10$), which corresponds to temperatures well below miliKelvin (see more details in \ref{App:ResultsSingle}). 
Many techniques have been used to cool the vibrational modes of a mechanical resonator, such as conventional cryogenic refrigeration processes or laser cooling techniques. In ref.~\cite{PRL113_027404_2014}, the authors showed that it is possible to cool the thermal motion of graphene down to a few tens of mK, in high-$Q$ microwave cavity.  Our cooling scheme could be implemented in combination with these other optomechanical cooling protocols; the phonon assisted cooling via vacuum interactions would then be used to cool the membrane further to the ground state. Based on that, we from now on we will choose a different set of initial graphene temperatures, $T=0.01, \,0.1$~K for our numerical simulations.
If we fix the value of the coupling strength $g$ and initial temperature, we observe that the minimum value of the cooling occurs for $\vartheta_{(1)} \pc{= \omega +  2 g^2 \md{\alpha}^2 / \nu } \simeq \nu$ . In fact, optimal cooling occurs in a narrow interval around $\vartheta_{(1)}/ \nu =1$ (see the left panel of figure \ref{Fig:1mode} for illustration). The eigenvalues of $A$ determine the relaxation time, which is given by the inverse of that having the smallest real part. For $g=0$, the relaxation time is given by the mechanical relaxation time $\kappa^{-1}$ $\pc{\tau \sim 0.2~\text{s}}$. Instead, for $g \neq 0$, we can obtain much larger decay rates (see the right panel of figure \ref{Fig:1mode} for illustration).

\subsubsection*{Acoustomechanical entanglement in a single mode membrane}

In order to establish the conditions under which the flexural and the phononic modes are entangled, we consider the logarithmic negativity $E_\mathcal{N}$ \cite{PRA65_032314_2002,PRA70_022318_2004} defined as
\begin{eqnarray}
E_\mathcal{N} = \max \pr{0, -\ln 2 \eta^- \pc{\mathcal{V}_\text{bip}}},
\end{eqnarray}
where $\mathcal{V}_\text{bip}$ is a generic $4\times 4$ correlation matrix associated with the bipartite system 
\begin{eqnarray}
\mathcal{V}_\text{bip} = \begin{pmatrix}
\mathcal{A} & \mathcal{C} \\
\mathcal{C}^T & \mathcal{B}
\end{pmatrix},
\end{eqnarray}
and $\eta^- \pc{\mathcal{V}_\text{bip}}$ is given by
\begin{eqnarray}
\eta^- \pc{\mathcal{V}_\text{bip}} \equiv \frac{1}{\sqrt{2}} \sqrt{\Sigma\pc{\mathcal{V}_\text{bip}} -\sqrt{\Sigma\pc{V_\text{bip}}^2 - 4 \det \mathcal{V}_\text{bip}}  }
\end{eqnarray}
with 
\begin{eqnarray}
\Sigma\pc{\mathcal{V}_\text{bip}} \equiv \det \mathcal{A} + \det \mathcal{B}- 2 \det \mathcal{C}.
\end{eqnarray}
A Gaussian state is entangled if and only if $\eta^- \pc{\mathcal{V}_\text{bip}} < 1/2$, which is equivalent to the positive partial transpose criterion, a necessary and sufficient condition for Gaussian states \cite{PRL84_2726_2000}. For a single mechanical mode, $\mathcal{V}_\text{bip} \equiv \mathcal{V}$ defined by equation~\eqref{eq:Lyapunov}. 
figure~\ref{Fig:1modeEntanglement} shows $\eta^-$ and the logarithmic negativity $E_N$ versus the normalized detuning $\vartheta_{(1)}/\nu$ for two different temperatures and coupling strengths.
One can see that, in all the cases, there is acoustomechanical entanglement, and this entanglement increases around the resonance condition $\vartheta_{(1)} \sim 1$ (where we achieve optimal cooling). The acoustomechanical entanglement is sensitive to both changes in the coupling strength and temperature (actually, it is very fragile to temperature variations), decreasing significantly for $T=0.1$~K. Moreover, for the same temperature, one can conclude that smaller coupling parameters lead to higher entanglement strength even though ground-state cooling is not achieved. Ground-state cooling and large steady-state acoutomechanical entanglement, $E_N \approx 5$, is shown to be possible via Casimir-Polder interactions, in agreement with ref.~\cite{SciChinPhysMA57_2276_2014}.

\begin{figure}
\centering
\begin{minipage}{.5\textwidth}
  \centering
  \includegraphics[width=7.5cm]{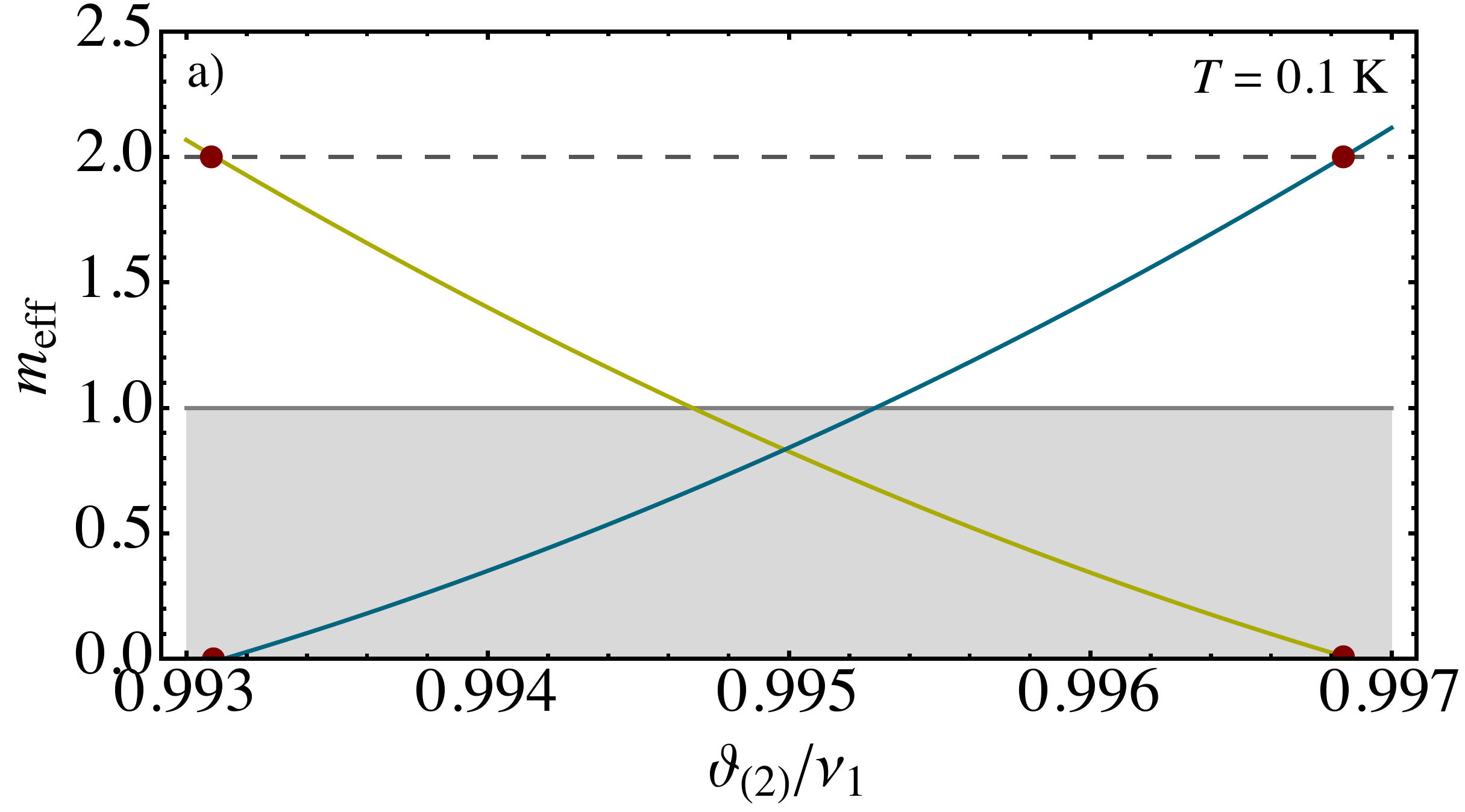}
\end{minipage}%
\begin{minipage}{.5\textwidth}
  \centering
  \includegraphics[width=7.5cm]{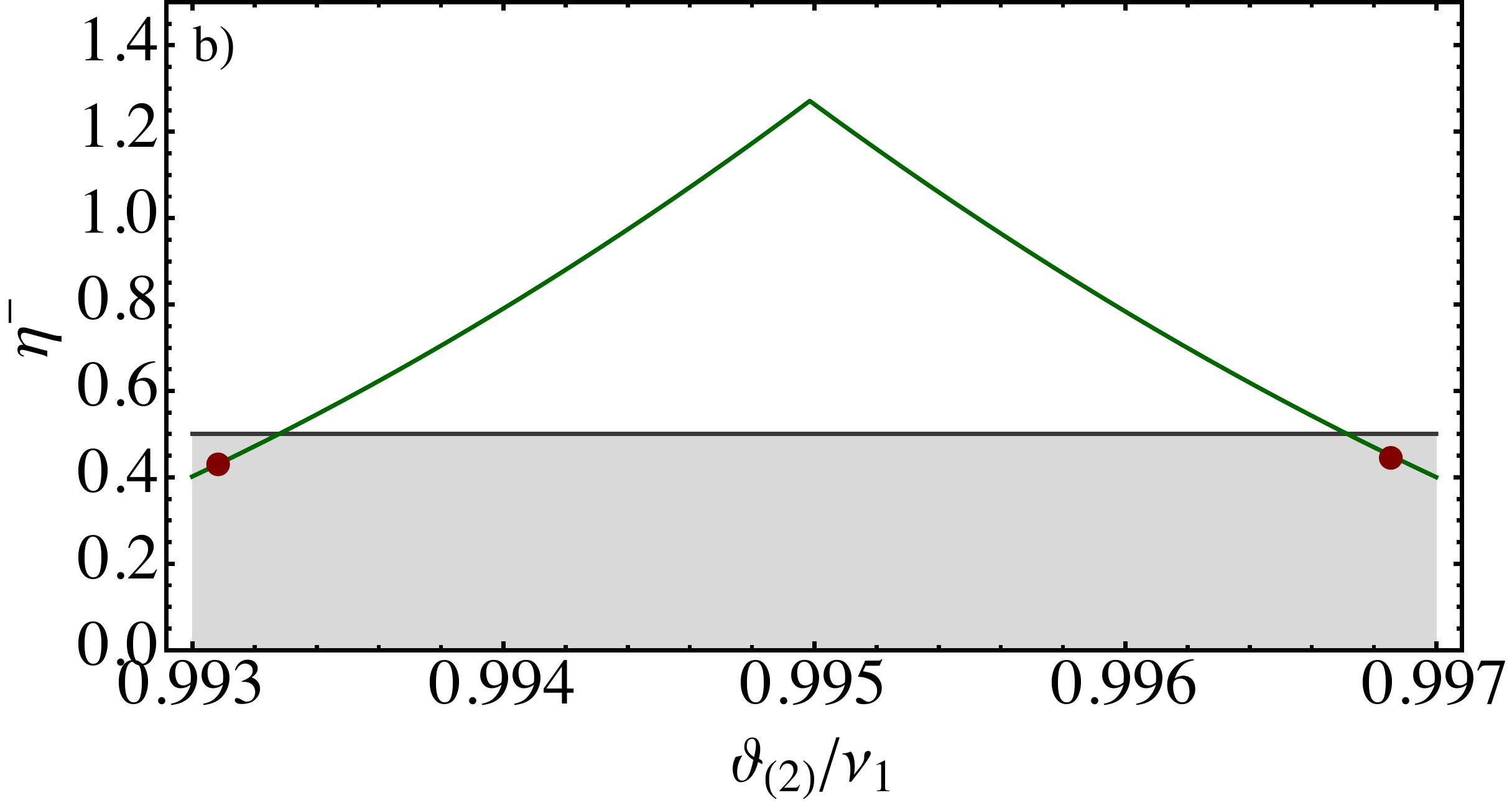}
\end{minipage} \\
\begin{minipage}{.5\textwidth}
  \centering
  \includegraphics[width=7.5cm]{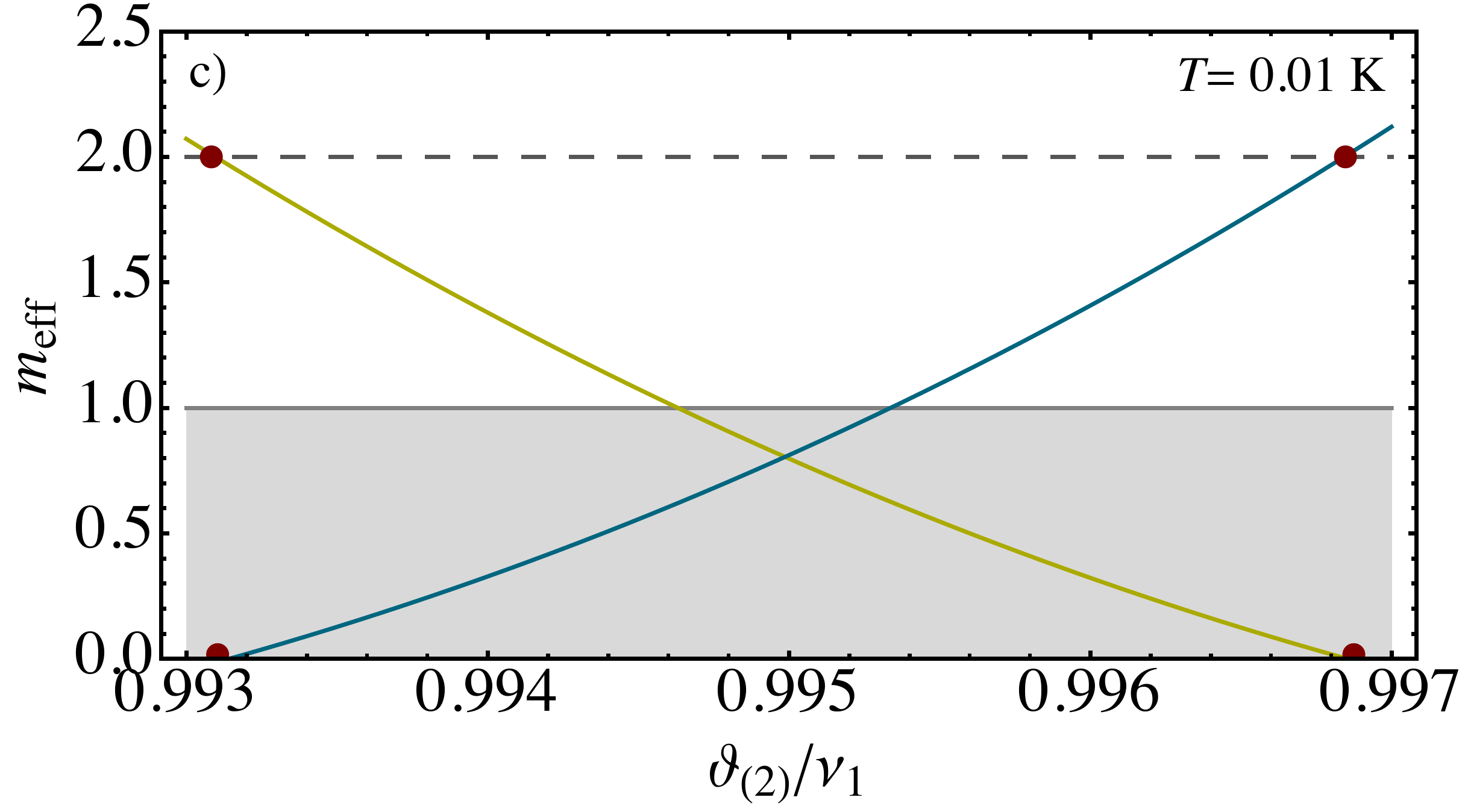}
\end{minipage}%
\begin{minipage}{.5\textwidth}
  \centering
  \includegraphics[width=7.5cm]{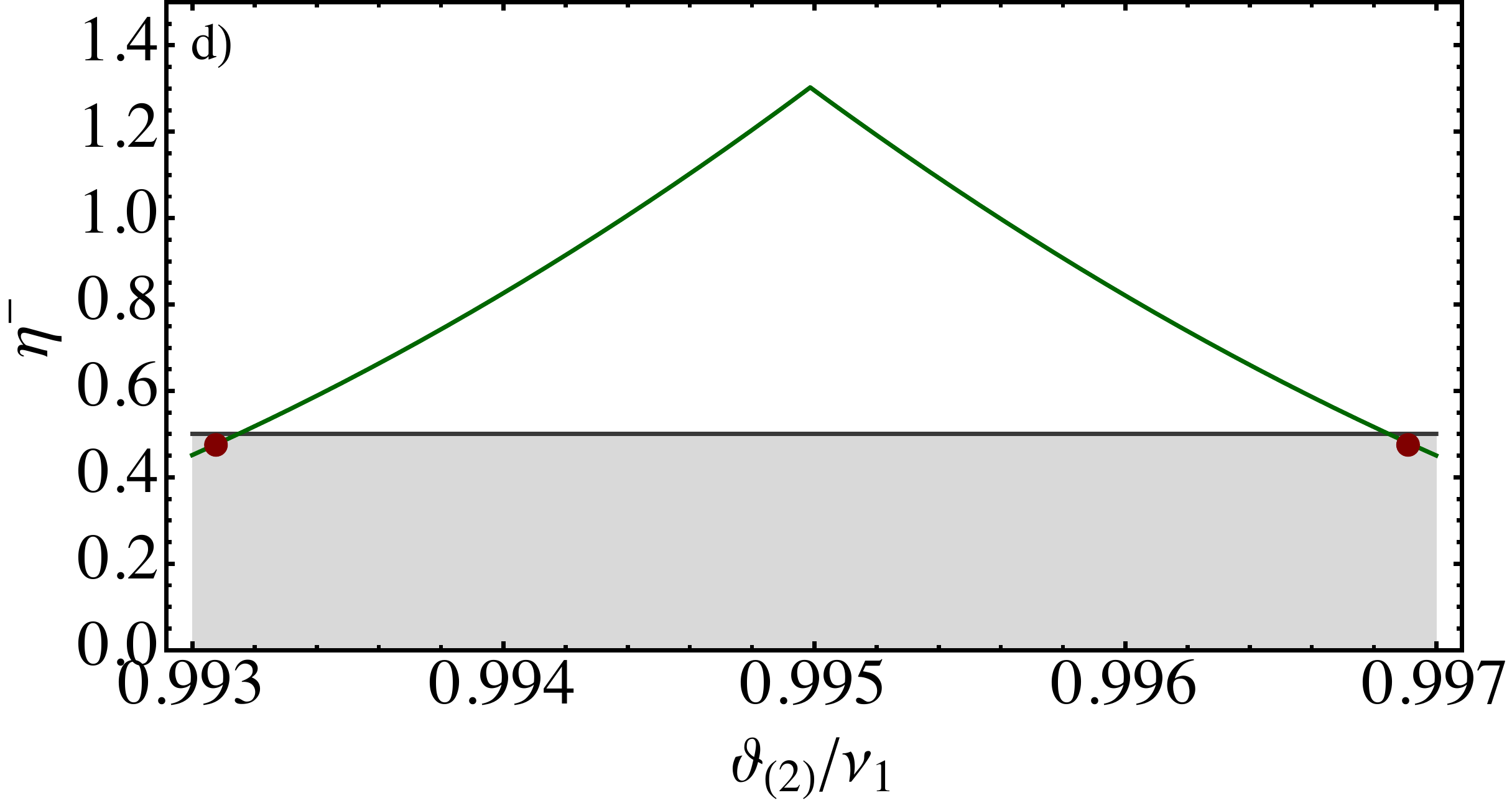}
\end{minipage}
\caption{(Colour online) a) and c) mean effective flexuron number of the modes $j=1$ (yellow) and $j=2$ (blue). b) and d) mechanical entanglement versus normalized detuning $\vartheta_{(2)}/\nu_1$.  The atomic and mechanical parameters are $\Gamma = 6.1$~MHz, $\omega_\text{ph}=477$~Hz, $\Delta = 45$~MHz, $\eta =0.15$, $\kappa =2$~Hz, $\nu_1 = 2$~MHz and $\nu_2=0.99 \nu_1$. On the top, for a) and b), $T=0.1$~K and $\Omega = 17.5$~MHz and $g_1 \approx g_2 \approx 43$~kHz; on the bottom, for c) and d), $T=0.01$~K, $\Omega=12$~MHz and $g_1 \approx g_2 \approx 40$~kHz. \label{Fig:2modesclose}}
\end{figure}

\section{Side-by-side membranes: Simultaneous cooling and entanglement properties}

We now consider two spatially separated side-by-side membranes placed near a single atomic cloud. There is no direct interaction between the membranes, but each membrane is coupled via Casimir-Polder forces to the atomic cloud. It is assumed that each membrane is restricted to a single flexural mode therefore, the dynamics of the system are therefore described by the drift matrix 
\begin{eqnarray}
A_{(2)} = \begin{pmatrix}
0 & - \nu_1 & 0 & 0 &0 & 0 \\
\nu_1 & -\kappa_1 & 0 & 0& - 2 \md{\alpha} g_1 & 0 \\
0 & 0 & 0 & - \nu_2 &0 & 0 \\
0 & 0 & \nu_2 & -\kappa_2 &- 2 \md{\alpha} g_2 & 0 \\
0& 0& 0 & 0 & -\gamma & \omega \\
- 2 \md{\alpha} g_1 & 0&- 2 \md{\alpha} g_2  &0 & -\omega & - \gamma
\end{pmatrix}.
\label{eq:A(2)}
\end{eqnarray}

There are two distinct situations, depending on the difference between the two flexural frequencies. If they are very different, the cooling of the mode (1) is not perturbed by the presence of the mode (2), like the case of figure~\ref{Fig:2modes}. On the other hand, if the frequencies of the two modes are similar, both modes are simultaneously cooled close to their ground state. 
As so, we set $\nu_2 = \nu_1 + \delta$, with $\delta$ being small. For our numerical results, we considered a difference of 1\% between the two frequencies. The two modes are optimally cooled at two well-distinct values of $\vartheta_{(2)}$ and one can efficiently cool both modes if one fixes the detuning within a very narrow interval halfway between the two mechanical resonances,$\vartheta_{(2)} \approx \pc{\nu_2+ \nu_1} /2$.  The value of $\gamma$ can be tuned by changing $\Omega$ and, as a result, one can tune the value of $m_\text{eff}$ to be lower than 1, thus entering in the quantum regime (see figure~\ref{Fig:2modesclose}). 

For this case, we would also like to study the steady-state acoustomechanical and mechanical entanglement.
The steady-state correlation matrix $\mathcal{V}$ for two mechanical modes is a $6 \times 6$ matrix which can be written in terms of blocks of $2\times 2$ matrices as
\begin{eqnarray}
\mathcal{V} = \begin{pmatrix}
\mathcal{A}_1 & \mathcal{C}_{12} & \mathcal{D}_1\\
\mathcal{C}_{12}^T & \mathcal{A}_2 & \mathcal{D}_2\\
\mathcal{D}^T_1 & \mathcal{D}^T_2 & \mathcal{B}
\end{pmatrix}.
\end{eqnarray}
To check if the two mechanical modes are entangled at the steady-state, one eliminates the entries in $\mathcal{V}$ that correspond to the atomic phonon field, to get
\begin{eqnarray}
\mathcal{V}_\text{bip} = \begin{pmatrix}
\mathcal{A}_1 & \mathcal{C}_{12}\\
\mathcal{C}_{12}^T & \mathcal{A}_2
\end{pmatrix}.
\end{eqnarray}
Alternatively, in the case we would like to analyse the entanglement between one of the mechanical modes and the phonon modes, it suffices to eliminate the rows and columns that correspond to the other mechanical mode from the matrix $\mathcal{V}$
\begin{eqnarray}
\mathcal{V}_\text{bip} = \begin{pmatrix}
\mathcal{A}_{1,2} & \mathcal{D}_{1,2}\\
\mathcal{D}_{1,2}^T & \mathcal{B}
\end{pmatrix}.
\end{eqnarray}

The entanglement properties of the mechanical steady-state of the system will again strongly depend on the experimental situations. Although these two modes are not directly interacting, they can become entangled at the steady-state via the Casimir-Polder interaction with the atomic cloud. As it has been previously seen, when the two modes are well separated, either simultaneous cooling of the membranes or (purely) mechanical entanglement is not achievable. The situation is drastically different when two mechanical modes become very close in frequency $\nu_2 \approx \nu_1$. For this situation, when $\vartheta_{(2)} \neq \pc{\nu_2+ \nu_1} /2$, although we do not have simultaneous cooling, the membranes may be entangled for a certain set of parameters. In figure~\ref{Fig:2modesclose}, we compare two cases for two different temperatures. In both cases, when we have optimal cooling of the membranes to $m_\text{eff}<1$, there is no mechanical entanglement between them. However, there might be other interesting states, such as the Fock states $\ket{n,0}, \, \ket{0, \, n}$, where both membranes can be mechanically entangled (we show, for instance, the particular case where $n=2$ for both temperatures). 
Moreover, although at first sight the results depicted in figure~\ref{Fig:2modesclose}, show that lower temperatures will give rise to weaker mechanical entanglement strengths, in fact, if compared in the same parameters range, we observe that entanglement is very fragile with respect to temperature. This means, that if we choose the same parameters as figure~\ref{Fig:2modesclose}~d), but a temperature sightly higher, for instance $T=0.02$~K, the two graphene membranes are no longer entangled.

\begin{figure}
\centering
\begin{minipage}{.5\textwidth}
  \centering
  \includegraphics[width=7.5cm]{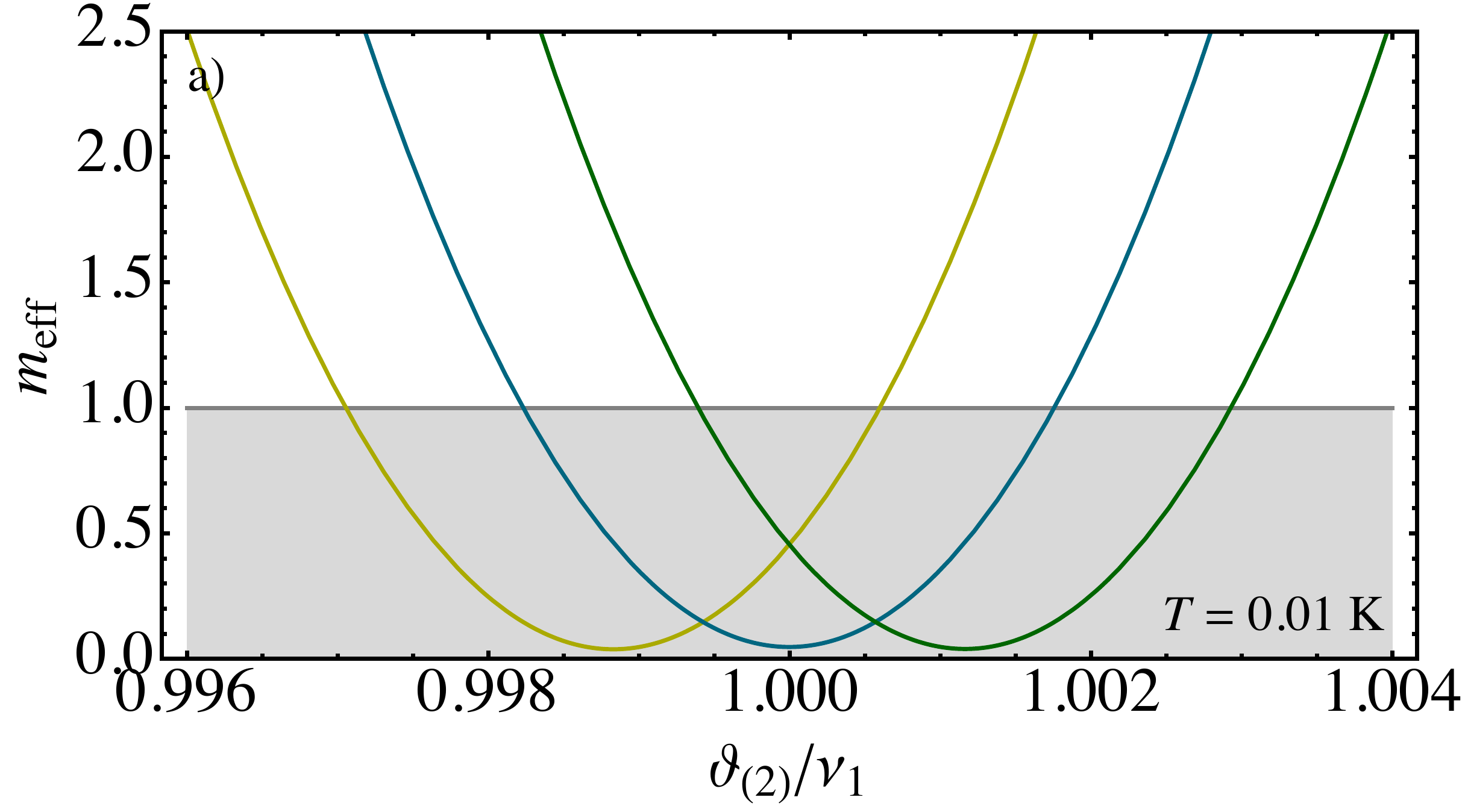}
\end{minipage}%
\begin{minipage}{.5\textwidth}
  \centering
  \includegraphics[width=7.5cm]{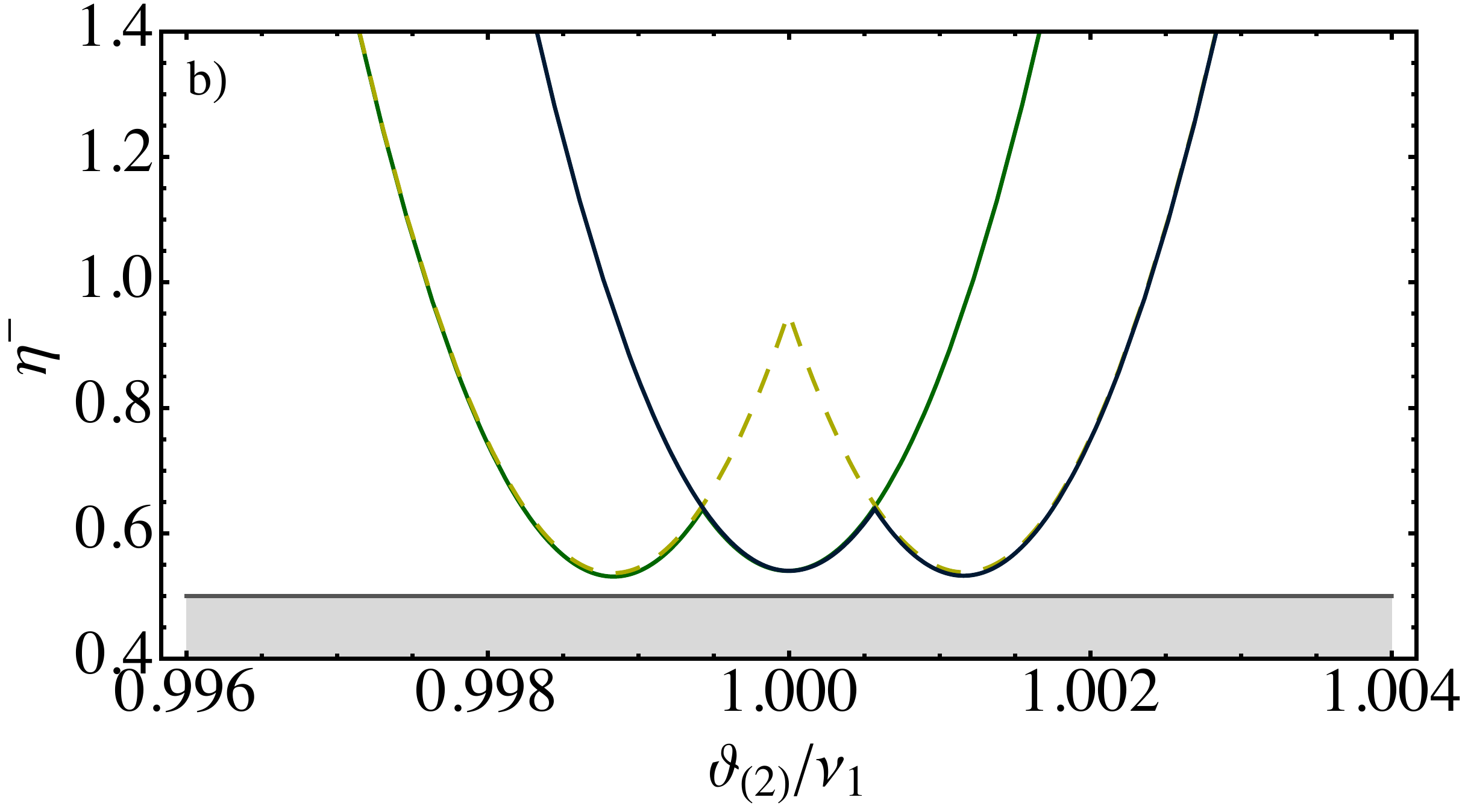}
\end{minipage}
 \caption{(Colour online) a) Mean effective flexuron number of the modes $j=1$ (yellow), 2 (blue), 3 (green) of a three side-by-side grahene sheets and b) mechanical entanglement between membrane 1 and 2 (green), 1 and 3 (dashed yellow) and 2 and 3 (blue) versus normalized detuning $\vartheta_{(3)}/ n_2$ for $T= 0.01$~K. The atomic and mechanical parameters are $\Gamma = 6.1$~MHz, $\omega_\text{ph}=477$~Hz, $\Omega = 17.5$~MHz, $\Delta = 45$~MHz, $\eta =0.15$, $\kappa =2$~Hz and $\nu_2 = 2$~MHz, $\nu_1=0.999 \nu_2$ and $\nu_3=1.001 \nu_2$ and $g_{1,2,3} \approx -4.8$~kHz.  \label{Fig:3modesT001}}
\end{figure}

\subsubsection*{Three modes: Simultaneous cooling}

Finally, we would like to analyse the case of three side-by-side membranes (see figure~\ref{Fig:setup}). In this situation, the drift matrix becomes a $8 \times 8$ matrix.
Repeating the procedures discussed above, we can compute the steady-state of the system. In order to have simultaneous cooling of the three membranes, we choose vibrational modes that are closer to each other than in the two-membrane case, and we set the 
initial temperature to $T=0.01$~K (see figure~\ref{Fig:3modesT001}). We have also studied the entanglement of the bipartite system composed by membrane 1 and 2 (green), 2 and 3 (dashed yellow) and 3 and 4 (blue). Although we find no bipartite entanglement between the membranes for a reasonable set of parameters, it is possible to cool all membranes down to the quantum regime $m_\text{eff}< 1$.

%%%%%%%%%%%%%%%%%%%%%%%%%%%%%%%%%%%%%%%%%%%%%%%%%%%%%%%%%%%%%%%%%%
\section{Conclusions \label{sec:conclusions}}

In this paper, we analysed the effect of the presence of one or more secondary modes in the sympathetic laser cooling of a graphene sheet coupled to an atomic cloud via Casimir-Polder interactions. We have seen that the simultaneous cooling of two modes crucially depends on the difference between their frequencies. We have shown that, for a single graphene sheet, the frequency of the fundamental and first excited flexural (out-of-plane) modes are too separated, such that the excited mode does not affect cooling. In fact, we observed that there are different experimental parameters that would allow us to cool one mode without affecting the other. 
Considering a multiple membrane system, where different non-interacting graphene sheets are placed side-by-side, we have shown that the modes are optimally cooled at well-distinct values of $\vartheta_{(N)}$ and one can efficiently cool both modes by setting the detuning within a very narrow interval halfway between the mechanical resonances.

Under the same conditions, we have also studied the acoustomechanical and mechanical entanglement considering multiple flexural modes. Large acoustomechanical entanglement can be achieved for single or multimode case, confirming previous results that indicated that vacuum forces enable steady-state acoustomechanical entanglement \cite{SciChinPhysMA57_2276_2014}. On the other hand, we demonstrated that the mechanical entanglement is very fragile and strongly depends on $\vartheta_{(N)}$. However, we are still able to prove that the mechanical states can become entangled thanks to the common interaction with the quantum gas.

\section*{Acknowledgements}
The authors acknowledge the Security of Quantum Information Group for the hospitality and for providing the working conditions. The authors would like to thank the support from Funda\c{c}\~{a}o para a Ci\^{e}ncia e a Tecnologia (Portugal), namely H.T. through scholarship SFRH/BPD/110059/2015 and S.R. via UID/EEA/50008/2013 project.

\appendix

\section{Linearisation procedure \label{App:Linearisation}}

When the ground-state cooling is achieved and if the system is stable, the system is characterized by a semiclassical steady-state where the phonons in the atomic system can be rewritten as a displacement transformation with an average amplitude $\alpha$ and a fluctuating part $\delta \hat{a}$.
The steady-state values of the position and momentum operators, as obtained from the stationarity of equation \eqref{eq:dotp} and \eqref{eq:dotq}, are given by
\begin{align}
p^s_j &= 0, \\
q_j^s &= \frac{\sqrt{2} g_j \md{\alpha}^2}{\nu_j}, \\
\alpha &= \frac{\xi}{i \vartheta_{(N)} + \gamma},
\label{eq:steadystate}
\end{align}
where we have defined $\vartheta_{(N)} = \omega + \sum_j 2 g_j^2 \md{\alpha}^2 / \nu_j$, with $N$ being the total number of modes considered. Then, we can linearise Eqs. \eqref{eq:dota}-\eqref{eq:dotp} around the steady-state values by setting $\hat{a}\to \alpha + \delta \hat{a}$, $\hat{q}_j \to q_j^s + \delta \hat{q}_j$ and $\hat{p}_j p_j^s \to \delta \hat{p}_j$. For small fluctuations, the second-order terms $\delta \hat{a}^\dagger \delta \hat{a}$ and $\delta \hat{a} \delta \hat{q}_j$ are ruled out of the dynamics. Introducing the phonon quadratures
\begin{align}
\delta \hat{X} &= \frac{\alpha \delta \hat{a}^\dagger + \alpha^* \delta \hat{a}}{\sqrt{2 \md{\alpha}^2}}, \\
\delta \hat{Y} &= \frac{i\pc{\alpha \delta \hat{a}^\dagger-\alpha^* \delta \hat{a} }}{\sqrt{2 \md{\alpha}^2}},
\end{align}
with $\pr{\delta \hat{X}, \delta \hat{Y} } = i$, we arrive at a system of linearised quantum Langevin equations
\begin{align}
\delta \dot{\hat{q}}_j &= - \nu_j \delta \hat{p}_j, \\
\delta \dot{\hat{p}}_j &=  \nu_j  \delta \hat{q}_j - \kappa_j  \delta \hat{p}_j - 2 g_j \md{\alpha} \delta \hat{X} + \zeta, \\
\delta \dot{\hat{X}} &=   \vartheta_{(N)}\delta \hat{Y} -  \gamma \delta \hat{X},\\
\delta \dot{\hat{Y}} &=  -  \vartheta_{(N)}\delta \hat{X} -  \gamma \delta \hat{Y} + \sum_j 2 \md{\alpha} g_j \delta \hat{q}_j.
\end{align}
These can be written in a compact form as
\begin{eqnarray}
\dot{u} (t) = A u(t) + n(t),
\label{eq:dotu}
\end{eqnarray}
where we defined the fluctuation vector
\begin{eqnarray}
u (t) &= \left( \delta \hat{q}_1 (t), \delta \hat{p}_1 (t),\dots, \delta \hat{q}_j (t),\delta \hat{p}_j (t),  \dots, \delta \hat{X} (t),\delta \hat{Y} (t) \right)^T,
\end{eqnarray}
the noise vector
\begin{eqnarray}
n (t) &= \left( 0, \zeta_1 (t) ,\dots,0, \zeta_j (t) , \dots   ,  0, 0 \right)^T,
\end{eqnarray}
and $A$ is the drift matrix that governs the dynamics of the expectation values. The solution of equation~\eqref{eq:dotu} is given by
\begin{eqnarray}
u (t) = M (t) u(0) + \int_0^t d \tau M(\tau) n (t-\tau),
\end{eqnarray}
where $M(t) = \exp \pc{A t}$. The system is stable and reaches its steady-state when all of the eigenvalues of $A$ have negative real parts, such that $M(\infty)=0$. Since the dynamics is linearised, the quantum steady-state of fluctuations is a zero-mean multipartite Gaussian state, fully characterized by its correlation matrix $\mathcal{V}$ whose elements read
\begin{eqnarray}
\mathcal{V}_{lm} = \frac{\av{u_l (\infty) u_m (\infty) + u_m (\infty) u_l (\infty)}}{2}.
\end{eqnarray}
When the system is stable, one gets
\begin{eqnarray}
\mathcal{V}_{lm} = \sum_{k,l} \int_0^\infty \!\!\!\! d \tau \int_0^\infty \!\!\!\! d \tau' M_{ik} (\tau) M_{jl} (\tau') \Phi_{kl} (\tau - \tau'),
\end{eqnarray}
where $\Phi_{kl} (\tau - \tau') = \av{n_k(\tau) n_l (\tau') + n_l (\tau') n_k (\tau)}/2$ is the stationary noise correlation function. It is clear, as stated before, that the membrane Brownian noise $\zeta (t)$ is not delta-correlated and therefore does not describe a Markovian process \cite{PRA63_023812_2001,PRA77_033804_2008}. Quantum effects are achievable only for oscillators with a large mechanical quality factor $Q_j = \nu_j / \kappa_j \gg 1$, such as the case of graphene \cite{NatNano9_820_2014}. 
In ref.~\cite{JMathPhys6_504_1965,PRL46_1_1981}, it has been shown that if a process is purely Gaussian random and if we can treat the dynamical system quantum mechanically and interpret the canonical distribution of the heat bath also quantum mechanically, then, in this limit, one can recover a Markovian process and $\zeta (t)$ satisfies \cite{PRL98_030405_2007}
\begin{eqnarray}
\frac{\av{\zeta(\tau) \zeta (\tau') +\zeta (\tau') \zeta (\tau)}}{2} \approx \kappa_j \pc{2 m_j +1 } \delta (\tau-\tau').
\end{eqnarray}
Since the components of $n(t)$ are now uncorrelated, we get
\begin{eqnarray}
\Phi_{kl} (\tau - \tau') = D_{kl} \delta (\tau - \tau'),
\end{eqnarray}
where $D= \text{Diag}\pr{0,\kappa_j \pc{2 m_j +1 },0,0 }$ is the diagonal diffusion matrix determined by the noise correlation functions. Thus, we find
\begin{eqnarray}
\mathcal{V} = \int_0^\infty d \tau  M(\tau) D M^T(\tau).
\end{eqnarray}
When the stability condition $M(\infty)=0$ is satisfied, one gets the following equation for the steady-state correlation matrix
\begin{eqnarray}
A \mathcal{V}+ \mathcal{V} A^T = -D. 
\end{eqnarray}

\section{Results single mode \label{App:ResultsSingle}}

\begin{figure}
\centering
\begin{minipage}{.5\textwidth}
  \centering
  \includegraphics[width=7.5cm]{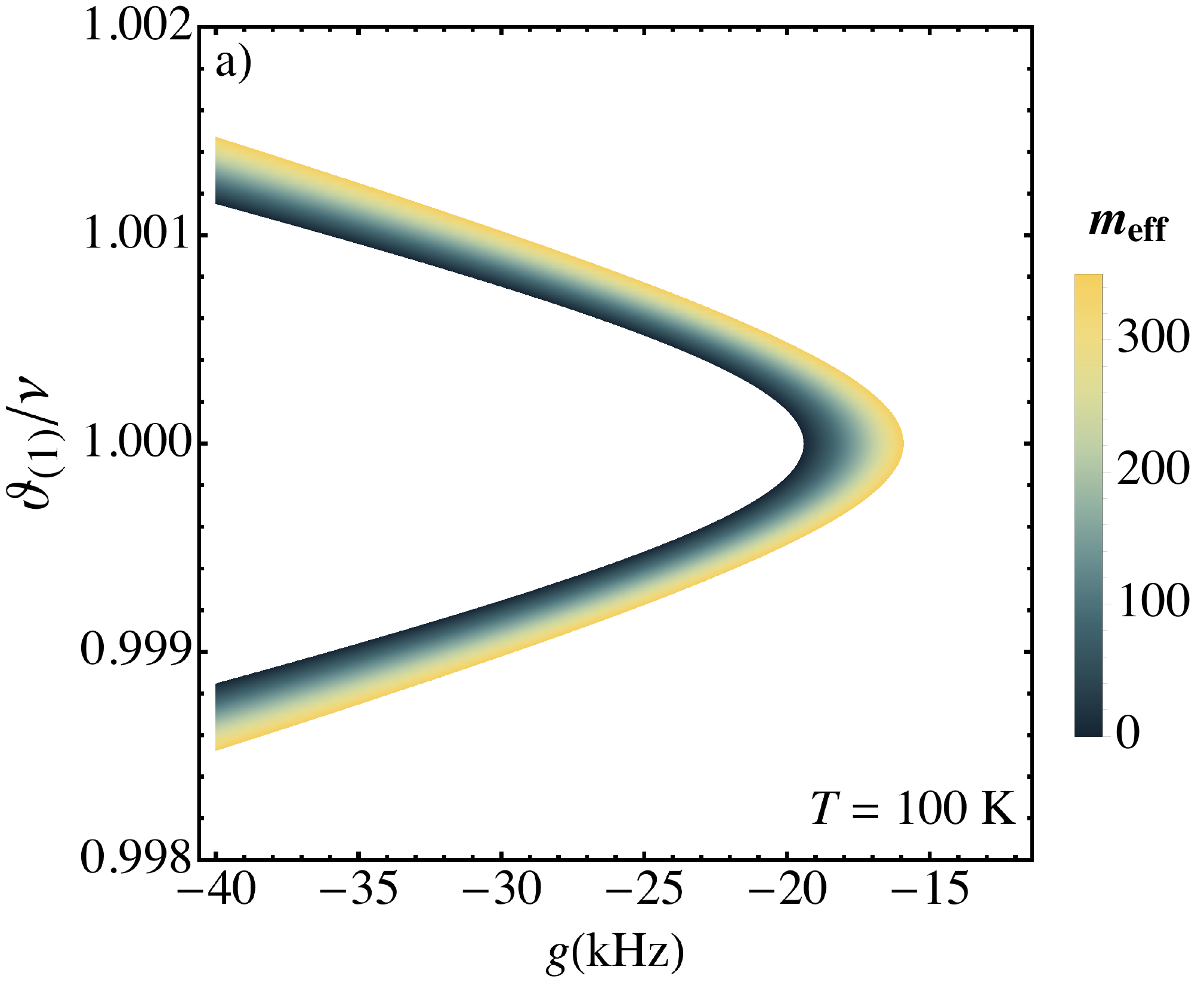}
\end{minipage}%
\begin{minipage}{.5\textwidth}
  \centering
  \includegraphics[width=7.5cm]{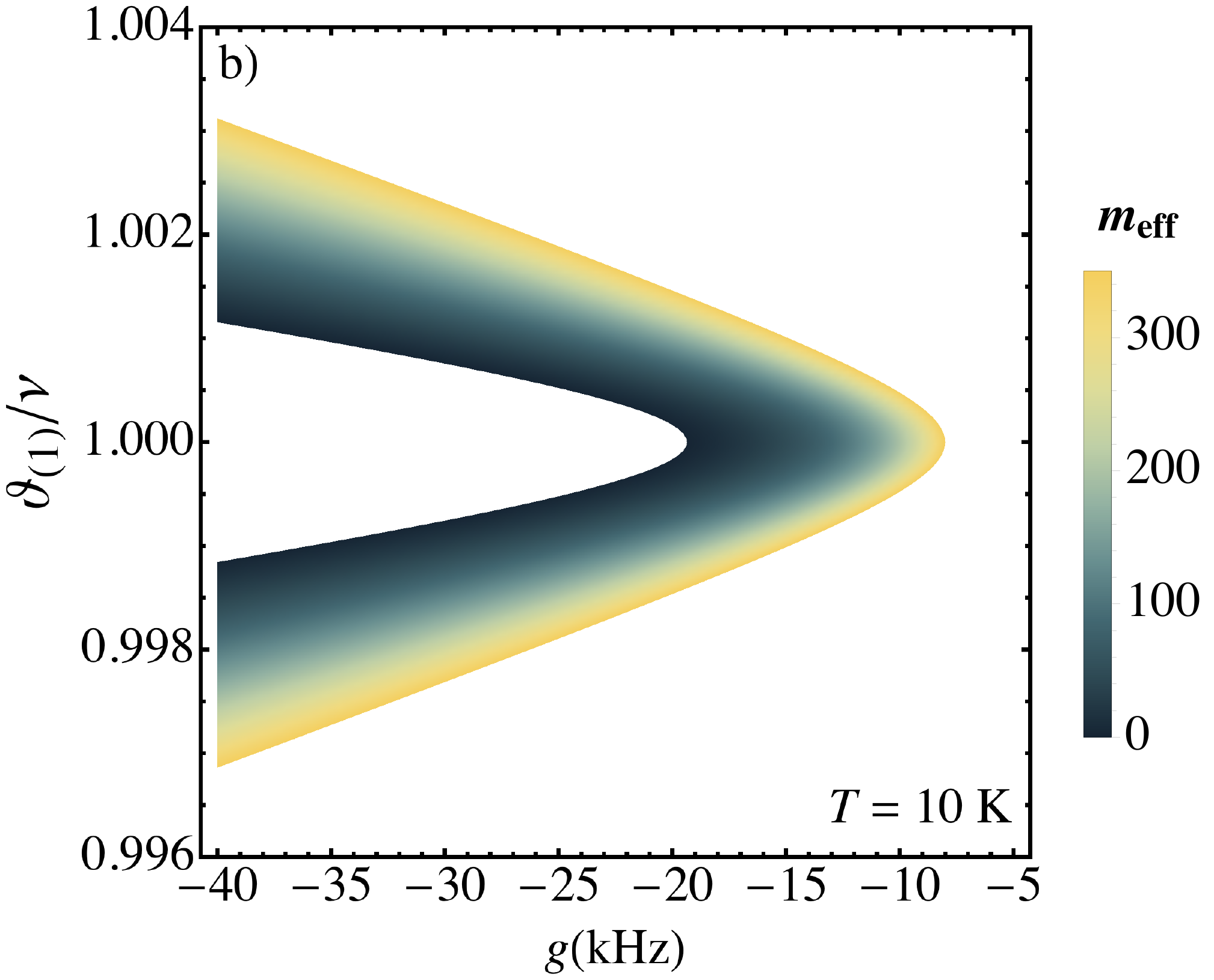}
\end{minipage}\\
\begin{minipage}{.5\textwidth}
  \centering
  \includegraphics[width=7.5cm]{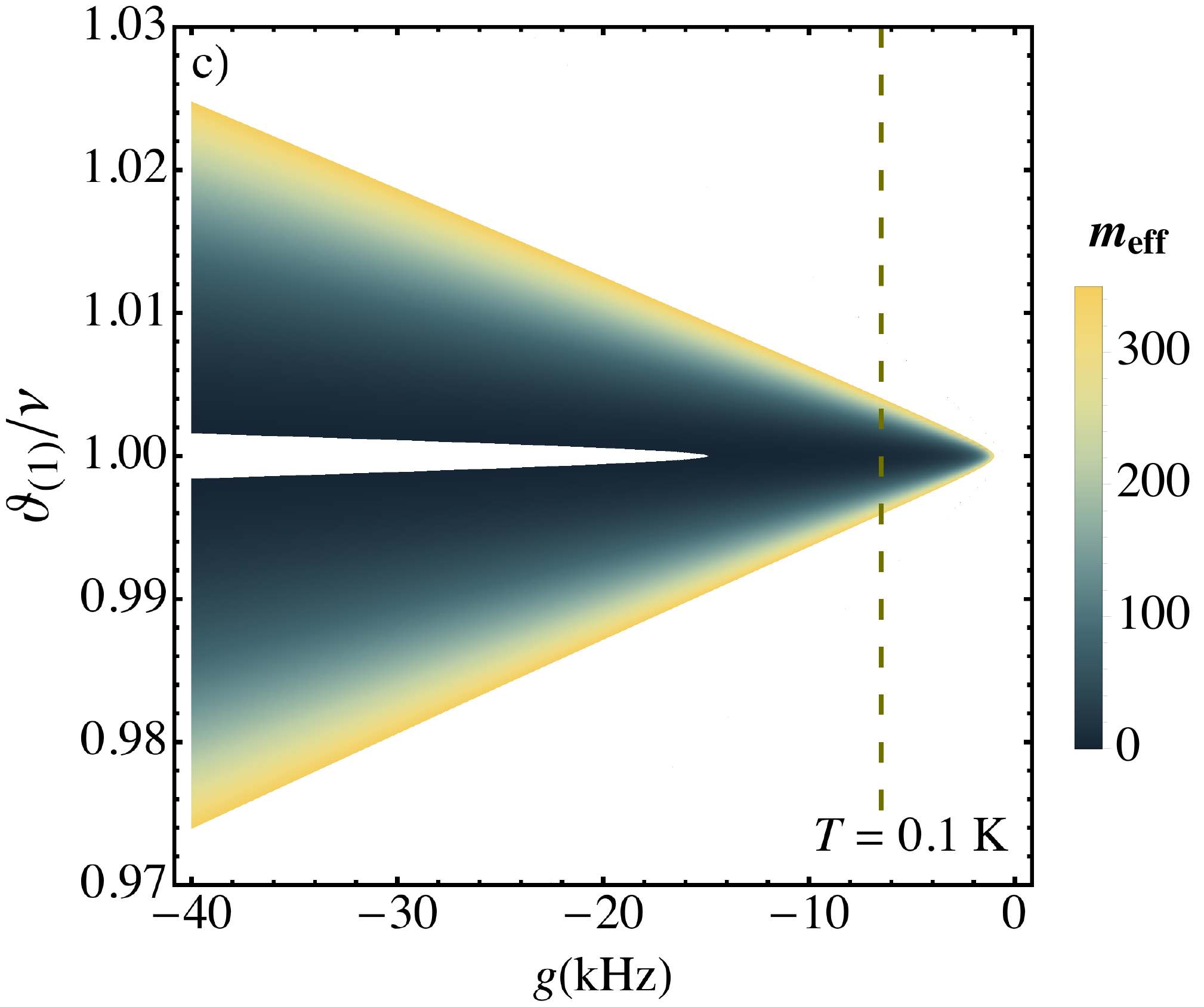}
\end{minipage}%
\begin{minipage}{.5\textwidth}
  \centering
  \includegraphics[width=7.5cm]{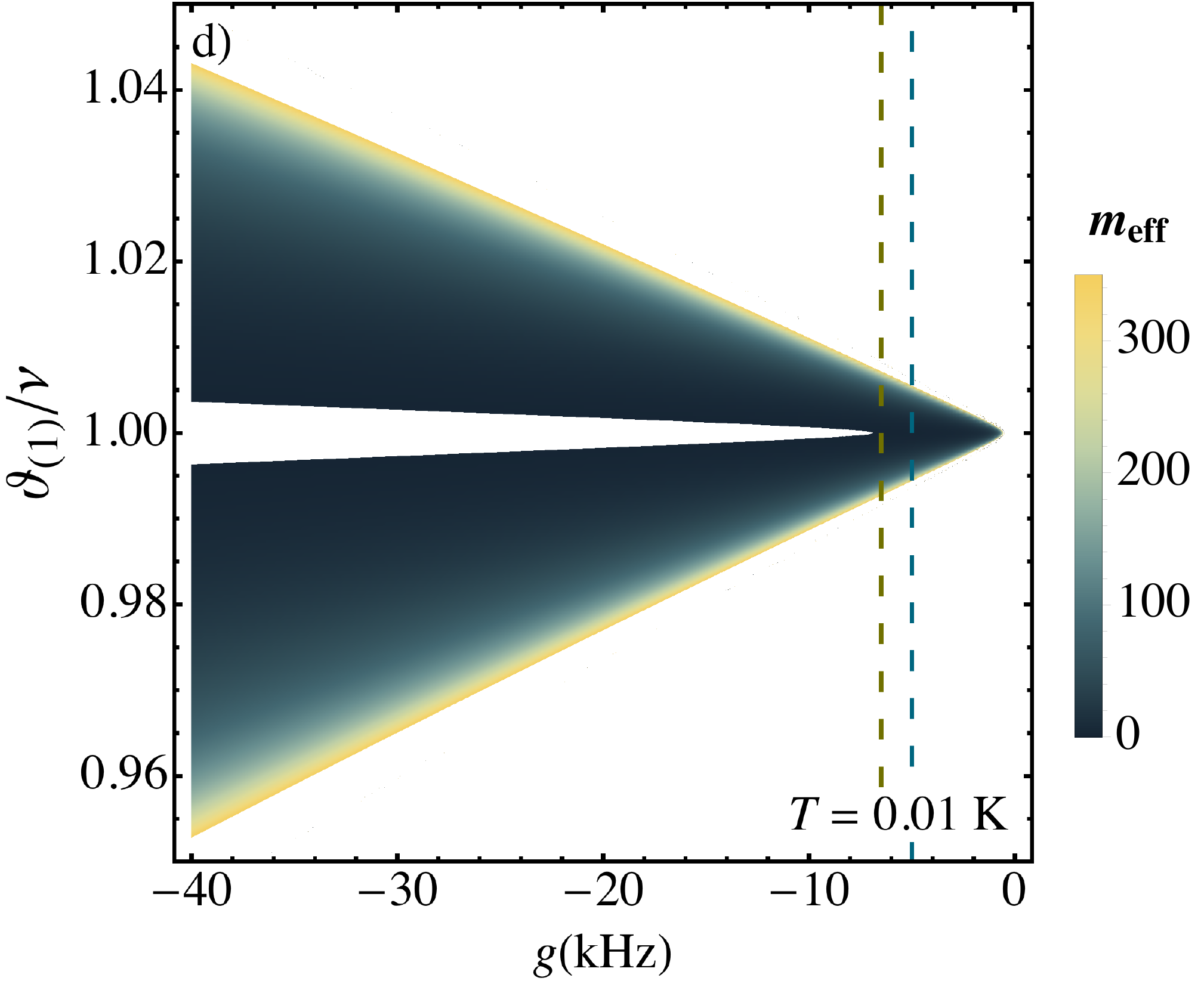}
\end{minipage}
  \caption{(Colour online) Density plot of the stationary state flexural mode number $m_\text{eff}$ of a single graphene sheet with a single vibrational mode versus normalized detuning $\vartheta_{(1)}/\nu$ and coupling parameter $g$. The atomic and mechanical parameters are $\Gamma = 6.1$~MHz, $\omega_\text{ph}=477$~Hz, $\Omega = 12$~MHz, $\Delta = 45$~MHz, $\eta =0.15$, $\kappa =2$~Hz and $\nu = 2$~MHz. 
a) corresponds to an initial temperature of $T=100$~K, b) to $T=10$~K, c) to $T=0.1$~K and d) to $T=0.01$~K. The dashed yellow lines correspond to a coupling strength $g=-6.5$~kHz and the dashed blue line to $g=-5$~kHz. \label{Fig:Cooling1mode}}
\end{figure}

To make a comparison with the results of reference~\cite{CoolingPaper}, in figure~\ref{Fig:Cooling1mode}, we have plotted $m_\text{eff}$ in terms of the tunable experimental parameters $\vartheta_{(1)}/\nu$ and coupling strength $g$. Since a more realist system must include losses in the vibrational motion of the membrane, the final $m_\text{eff}$ is temperature dependent, see differences between figure~\ref{Fig:Cooling1mode}a) for $T=100$~K, b) $T= 10$~K, c) $T=  0.1 $~K and d) $T=0.01$~K, which corresponds to initial flexuron occupation numbers of $10^6$, $10^5$, $10^3$ and $10^2$, respectively. We show that these results agree with our previous ones, that is, with an appropriate choice of $g$ and $\vartheta_{(1)}$, it is possible to have effective steady-states with a low number of flexural modes.

\subsubsection*{Two Flexural modes in a single membrane: Simultaneous cooling}

\begin{figure}
\centering
\begin{minipage}{.5\textwidth}
  \centering
  \includegraphics[width=7.5cm]{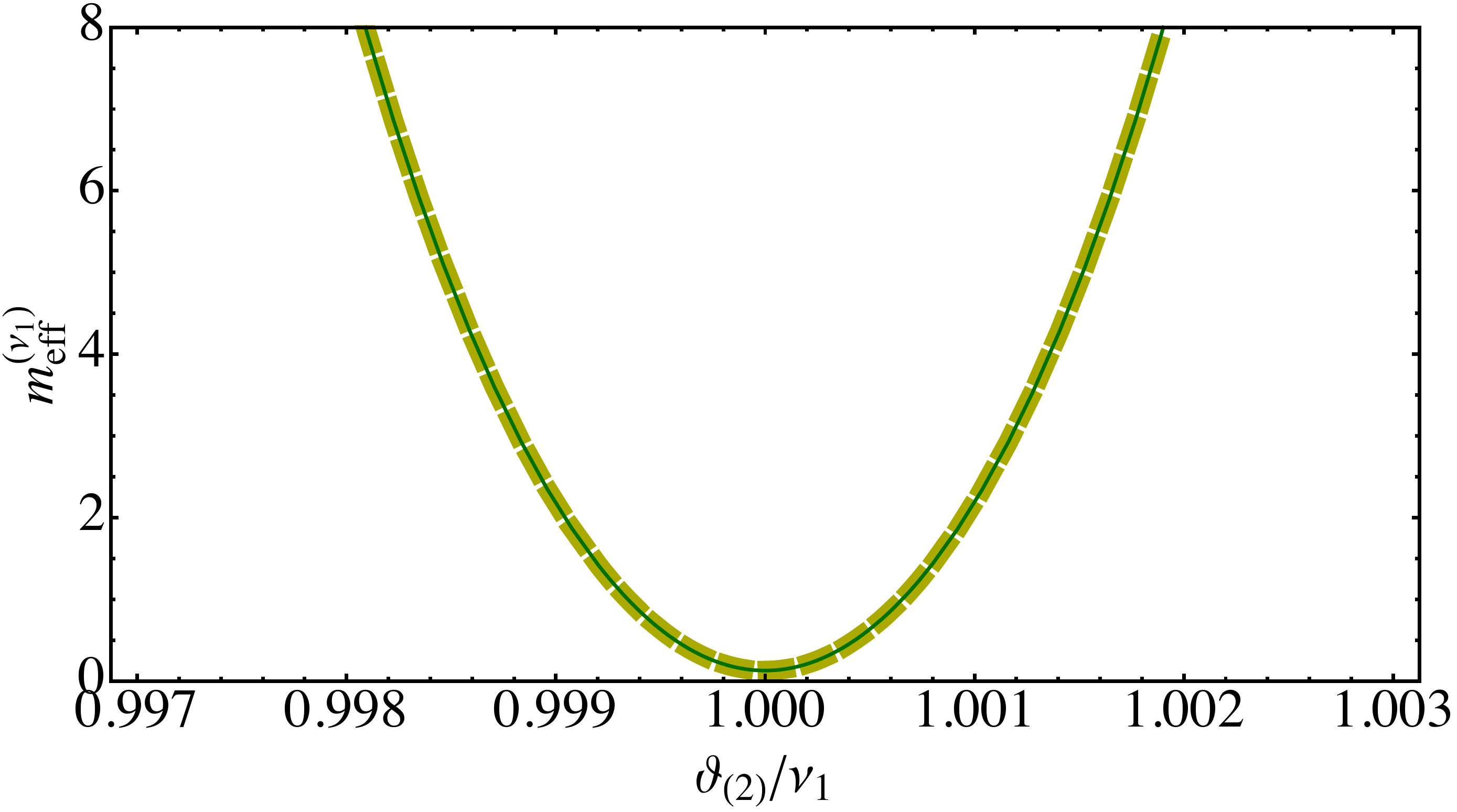}
\end{minipage}%
\begin{minipage}{.5\textwidth}
  \centering
  \includegraphics[width=7.5cm]{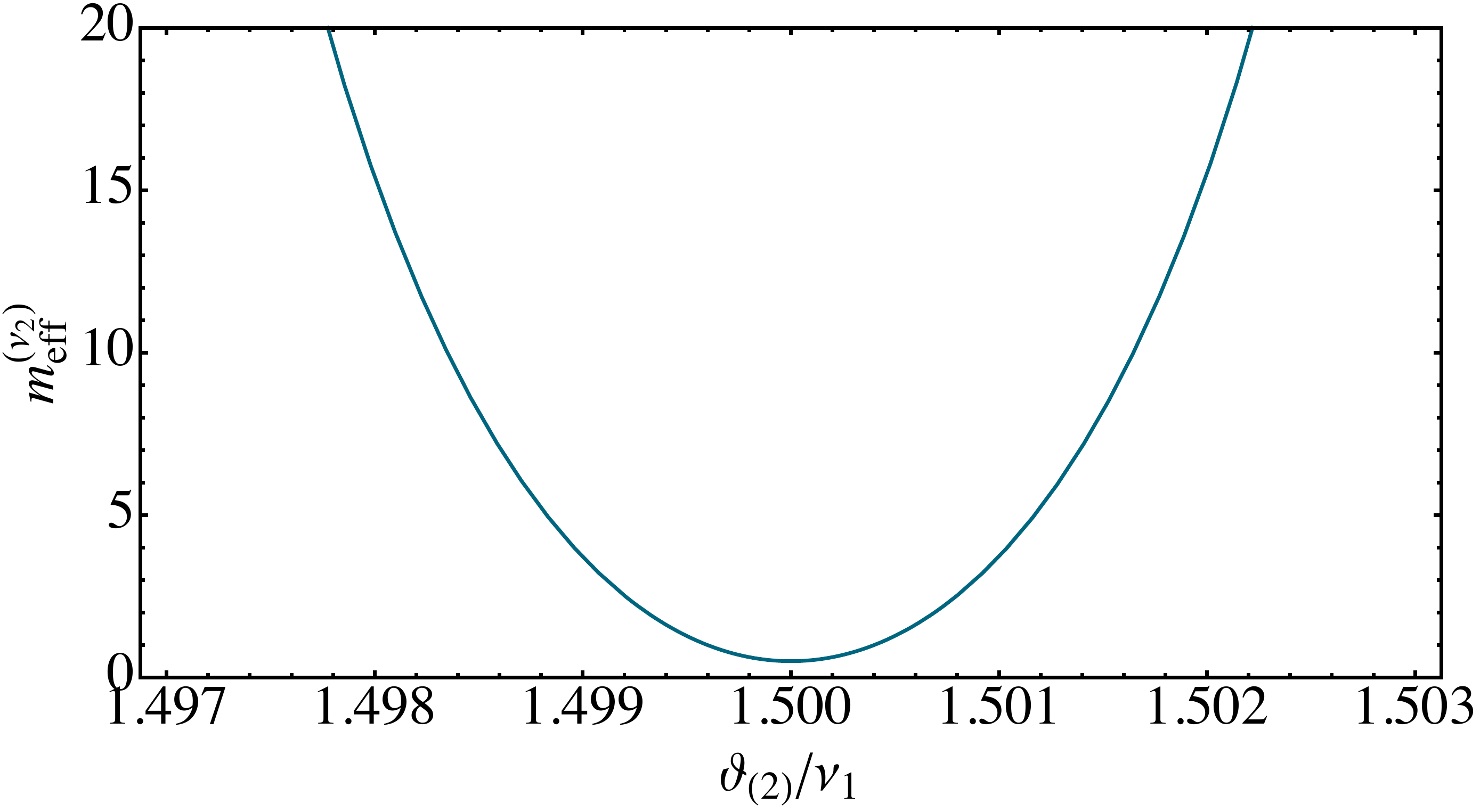}
\end{minipage}
  \caption{(Colour online) Mean effective flexuron number of the modes $j=1$ (left) and $j=2$ (right) of a  single graphene sheet versus normalized detuning $\vartheta_{(2)}/\nu_1$. The atomic and mechanical parameters are $\Gamma = 6.1$~MHz, $\omega_\text{ph}=477$~Hz, $\Omega = 12$~MHz, $\Delta = 45$~MHz, $\eta =0.15$, $\kappa =2$~Hz and $\nu_1 = 2$~MHz, $\nu_2=1.5 \nu_1$, with $T=0.01$~K and $g_1 \approx g_2 \approx -6.5$~kHz. The dashed yellow line on the top plot corresponds to the single mode case. \label{Fig:2modes}}
\end{figure}

In order to see if the presence of a secondary vibrational mode in a same membrane affects the ground-state cooling, we can exactly solve equation~\eqref{eq:Lyapunov} and analyse the stationary position and momentum variances of the two mechanical modes in order to calculate the effective flexuron number $m_\text{eff}^{(\nu_i)}$. To do so, we choose a parameter regime close to that of optimal cooling for a single mode. Our results indicate that when the two mechanical modes in a graphene sheet are well separated, $\nu_2 \simeq 1.5 \nu_1 $, the secondary mode does not disturb the cooling of the mechanical mode of interest, as we can see by the overlap of the curves for the single- and two-modes cases (see figure~\ref{Fig:2modes}). 
Furthermore, we observe that both modes are optimally cooled at two well distinct values of $\vartheta_{(2)}$: by cooling one, the other remains unaffected.

\begin{figure}
\centering
\begin{minipage}{.5\textwidth}
  \centering
  \includegraphics[width=7.5cm]{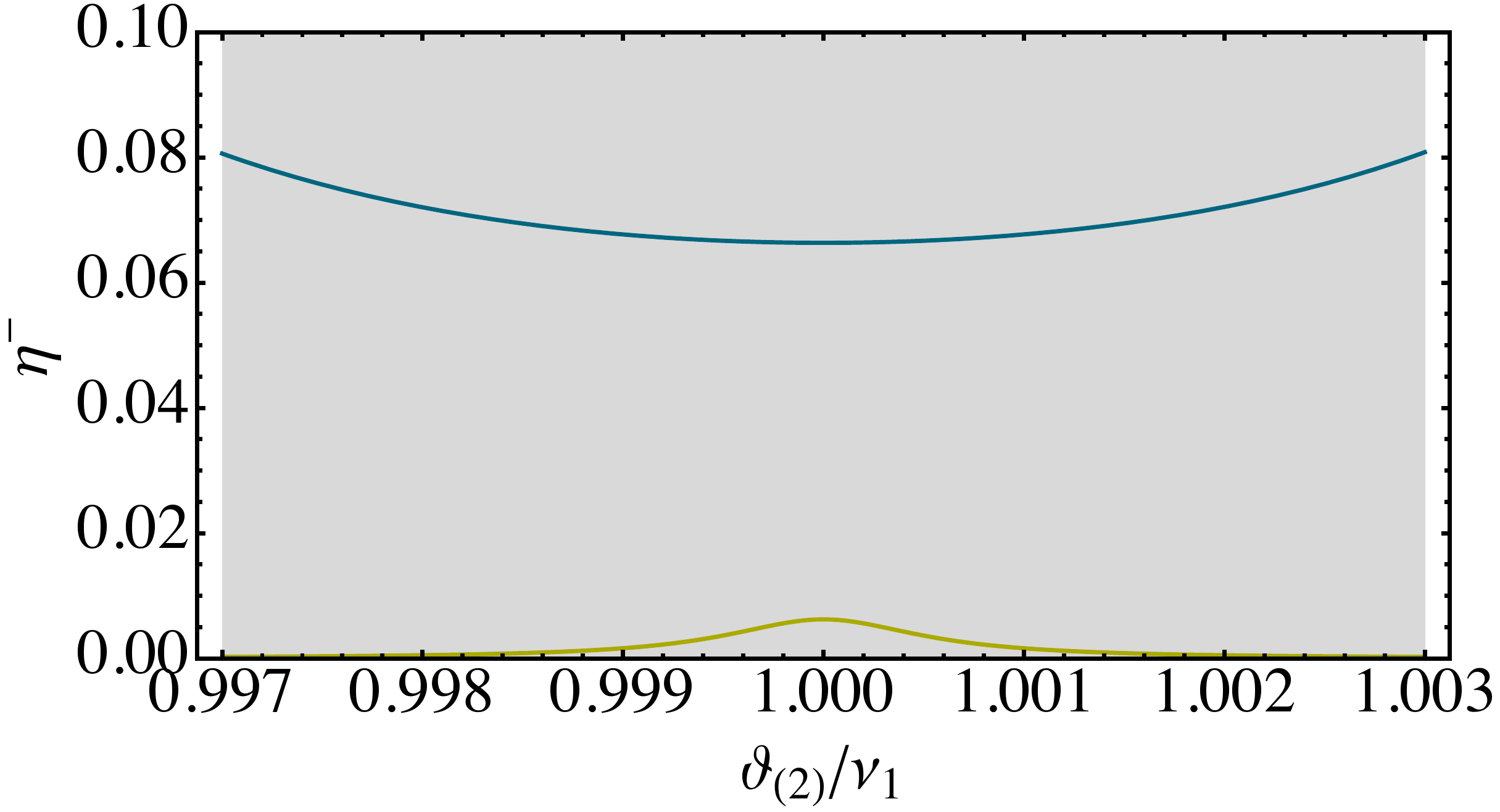}
\end{minipage}%
\begin{minipage}{.5\textwidth}
  \centering
  \includegraphics[width=7.5cm]{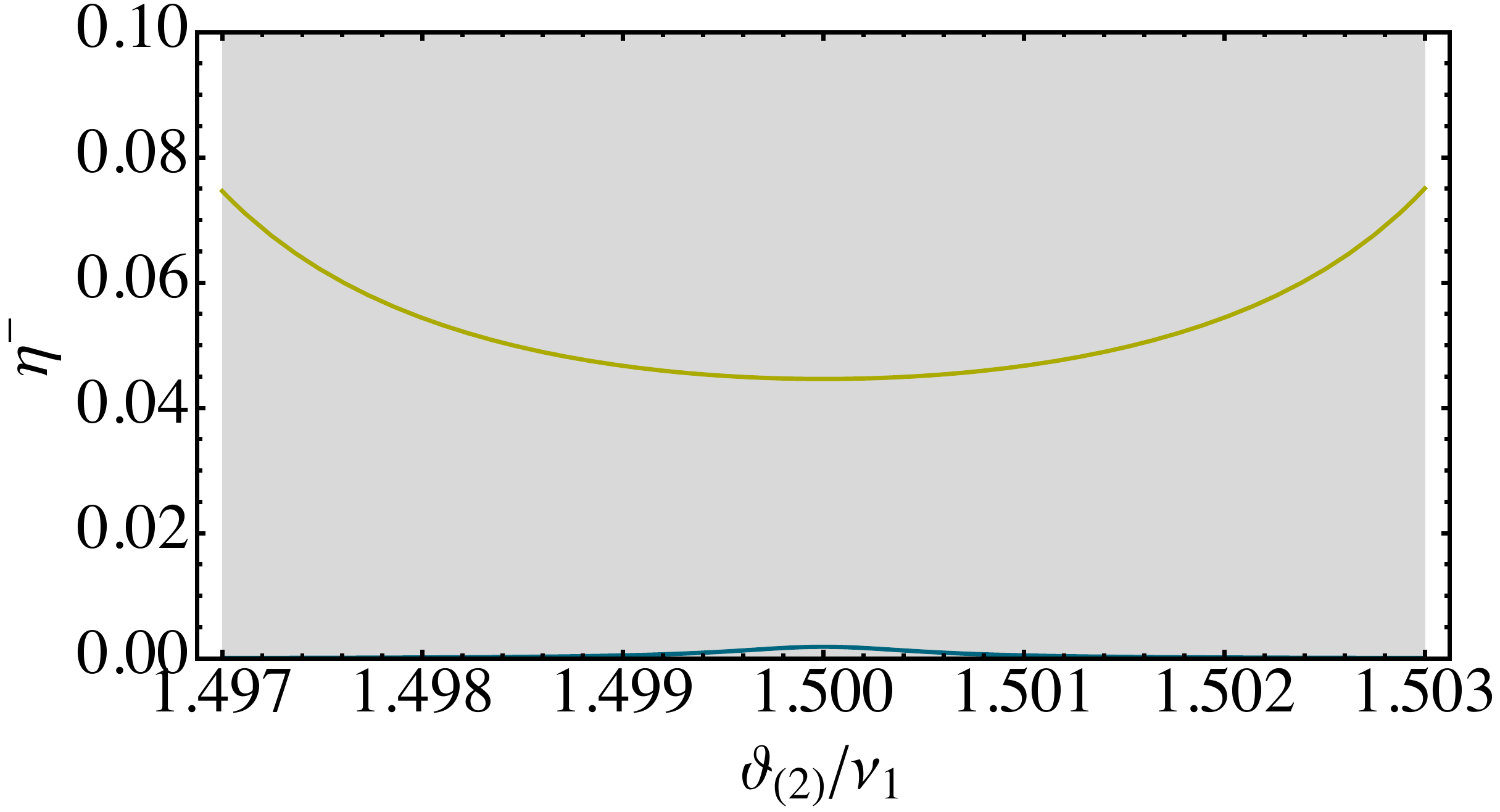}
\end{minipage}
  \caption{(Colour online) Acoustomechanical entanglement of the modes $j=1$ (yellow) and $j=2$ (blue) of a  single graphene sheet with the phononic mode versus normalized detuning $\vartheta_{(2)}/\nu_1$. The parameters are the same as in figure~\ref{Fig:2modes}. \label{Fig:2modesOpto}}
\end{figure}

\begin{figure}
\centering
\begin{minipage}{.5\textwidth}
  \centering
  \includegraphics[width=7.5cm]{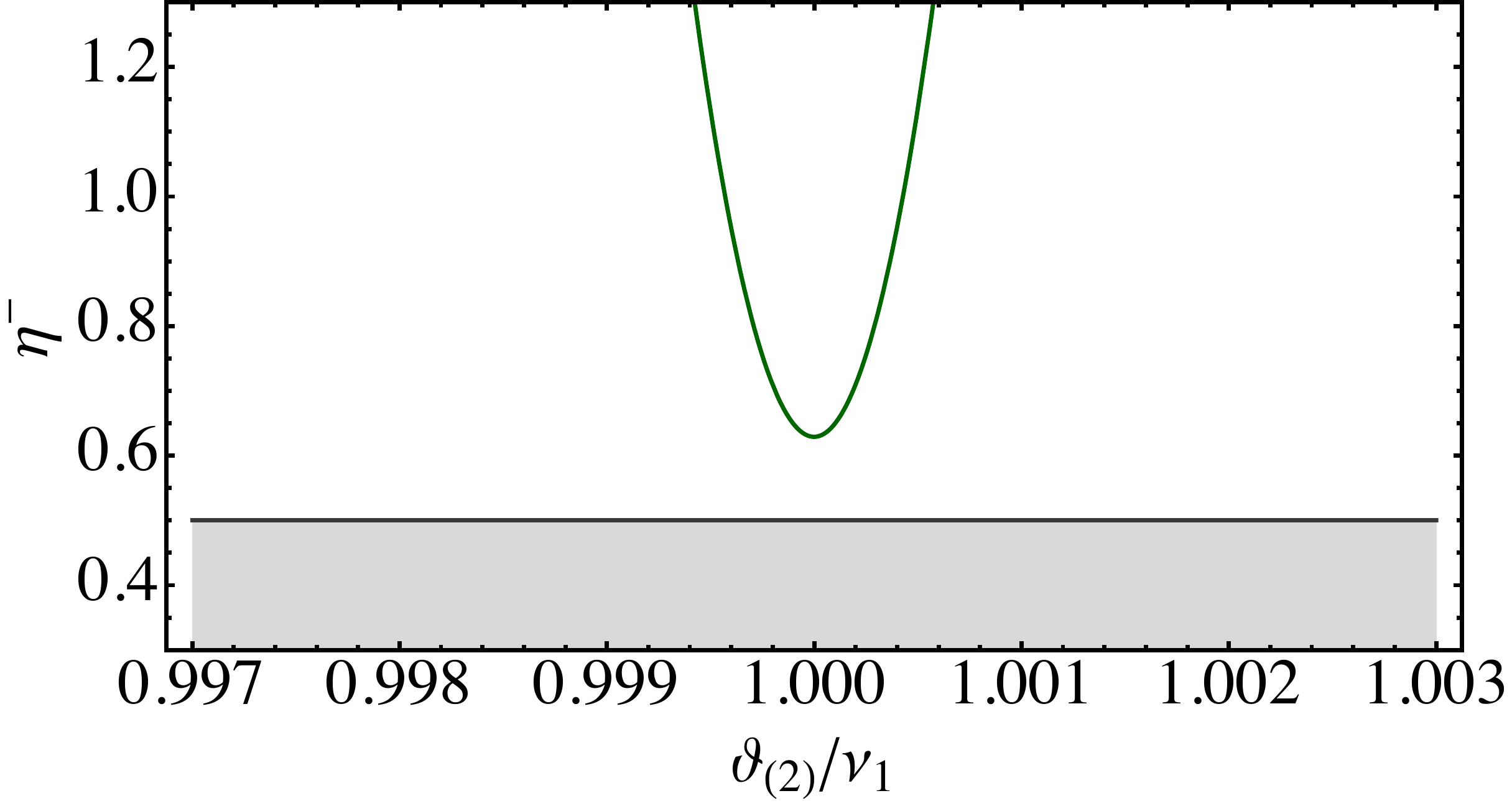}
\end{minipage}%
\begin{minipage}{.5\textwidth}
  \centering
  \includegraphics[width=7.5cm]{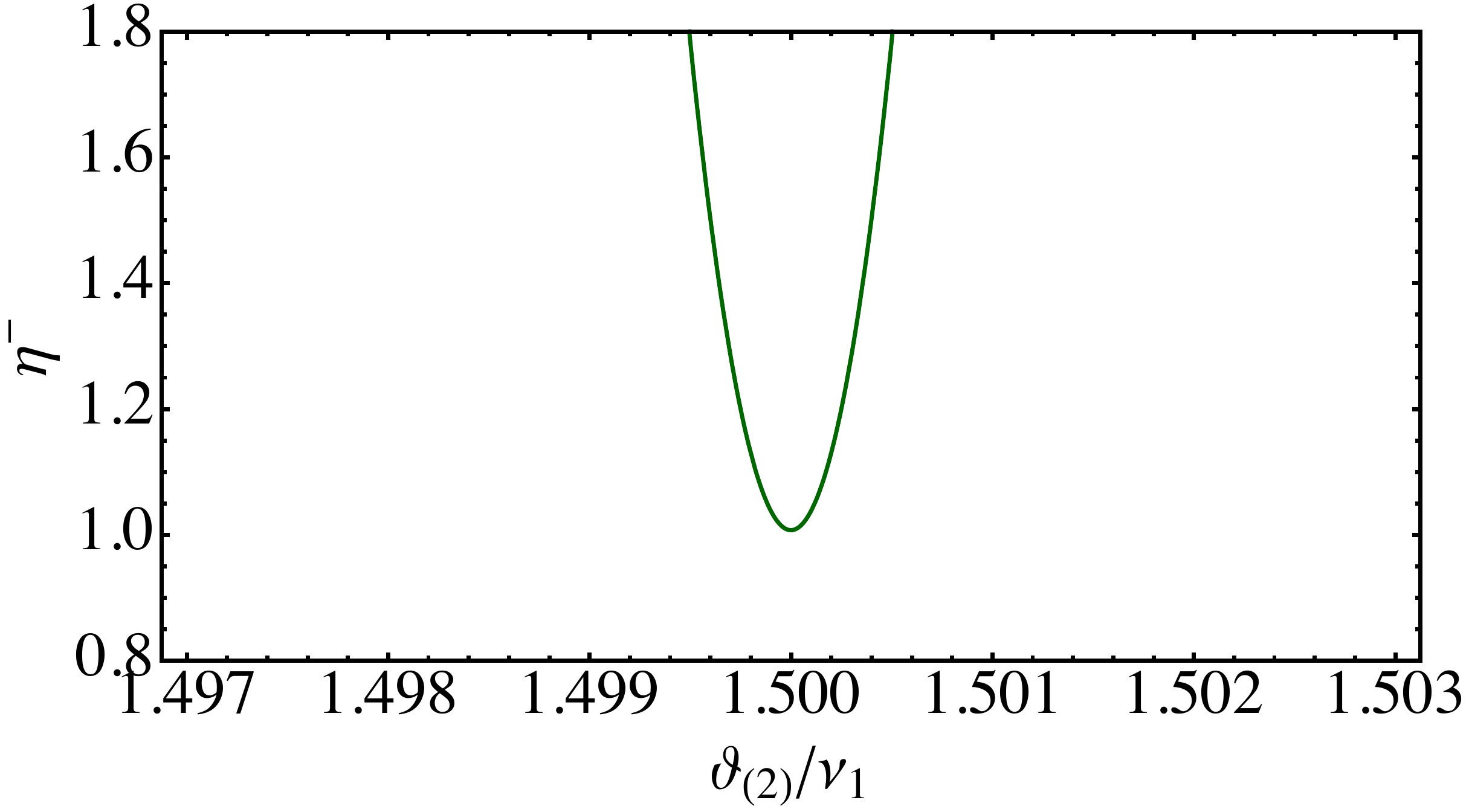}
\end{minipage}
  \caption{(Colour online) Mechanical entanglement of the modes $j=1$ and $j=2$ of a  single graphene sheet versus normalized detuning $\vartheta_{(2)}/\nu_1$. The parameters are the same as in figure~\ref{Fig:2modes}. \label{Fig:2modesMech}}
\end{figure}

We would also like to study the steady-state acoustomechanical and mechanical entanglement. Our results show that the presence of the second mode does not affect significantly the entanglement between the first mode and the phonon, see (yellow line) figure~\ref{Fig:2modesOpto}. Furthermore, we verify that the second mode is also entangled with the phononic mode (see blue line). However, we find that the purely mechanical entanglement between the first and second modes cannot be generated for this case, see figure~\ref{Fig:2modesMech}.


\end{document}